\documentclass{svproc}
\usepackage{graphicx}%
\usepackage{multirow}%
\usepackage{booktabs}%
\usepackage{amsmath,amssymb,amsfonts}%
\usepackage{mathrsfs}%
\usepackage{xcolor}%
\usepackage{textcomp}%
\usepackage{algorithm}%
\usepackage{listings}%
\usepackage[compatibility=false]{caption}
\usepackage{subcaption}
\usepackage{bm}
\usepackage{tikz}
\usetikzlibrary {arrows.meta,bending,positioning}
\usetikzlibrary{patterns}
\usepackage{matlab-prettifier}
\usetikzlibrary{shapes.multipart}
\usetikzlibrary{tikzmark}
\usepackage[utf8]{inputenc}
\usepackage{flushend}
\usepackage{tikz,pgfplots}
\usepackage{blkarray}
\usetikzlibrary{calc}
\usetikzlibrary{decorations.pathreplacing,angles,quotes}
\usetikzlibrary {decorations.fractals,spy}
\usepackage{tikz-dimline}
\usepackage{makecell}
\usepackage{mwe}
\newcolumntype{C}[1]{>{\centering\arraybackslash}m{#1}}

\newenvironment{rcases}
{\left.\begin{aligned}}
	{\end{aligned}\right\rbrace}

\pgfplotsset{compat=1.14}

\usepackage{url}

\begin{document}
\mainmatter 
\title{A Comparative Study on Robust Topology Optimization of Design-Dependent Pressure-Actuated Compliant Mechanisms with Quadrilateral Elements}
%
\titlerunning{Quad FE comparison of pressure actuated robust CM}  %
\author{Swagatam Islam Sarkar \and Prabhat Kumar\inst{*}}
\authorrunning{S.I. Sarkar and P. Kumar} 
%
\tocauthor{Swagatam Islam Sarkar and Prabhat Kumar}
%
\institute{Indian Institute of Technology Hyderabad, Telangana 502284, India \\ \inst{*}\email{pkumar@mae.iith.ac.in}}

\maketitle              
\begin{abstract}

This paper presents a comparative study of compliant mechanisms generated using a robust topology optimization technique involving design-dependent pressure loads. Design domains are parameterized using standard and higher-order quadrilateral elements. Both eroded and blueprint configurations are considered.
A min–max optimization model combined with an output-spring method is employed to extremize the mechanisms' output displacements. A volume and a strain energy constraint are applied to the blueprint and the eroded designs, respectively. The optimization process is executed using the method of moving asymptotes. Numerical experiments are performed to optimize the pressure-actuated inverter and gripper mechanisms using Q4, Q8, and Q9 elements, and the results are compared. The research highlights how quadrilateral element selection influences both the resulting topologies and performance characteristics.
\end{abstract}
\section{Introduction}\label{sec1}
A compliant mechanism (CM) comprises a single body with flexible parts, which is designed not only to provide sufficient flexibility to achieve the desired output but also to withstand applied external loads. Several approaches have been proposed for designing such mechanisms, including the pseudo-rigid-body model introduced by Howell et al.~\cite{howell1994method} and the topology optimization technique first presented by Ananthasuresh et al.~\cite{ananthasuresh1994strategies}. In this work, we employ the density-based TO technique and conduct a comparative study on the design of CMs using standard and higher-order quadrilateral finite elements under design-dependent loading conditions in the robust topology optimization framework.

Topology optimization (TO) is an engineering design technique that optimizes the distribution of material within a design domain. This technique provides lightweight, efficient, and sustainable designs that meet the desired objective. The finite element method is the most common approach to solving the related boundary value problems in the density-based TO framework. The design domain is typically parameterized by finite elements (FEs), and each element is given a design variable. Various FE types, such as triangular/quadrilateral/hexagonal/polygonal elements~\cite{saxena2007honeycomb,kumar2023honeytop90,kumar2015topology}, have been used to solve these TO problems. Bilinear four-node quadrilateral FEs (Q4) are widely employed to discretize the design domain, due to their simplicity, which they offer for the simulations~\cite{sarkar2025topology}. However, numerical instabilities, such as checkerboard patterns and point connections, are commonly observed in the optimized solutions using Q4 FEs~\cite{sigmund1998numerical}. Methods like density filtering~\cite{bruns2001topology}, robust formulation~\cite{sigmund2009manufacturing,wang2011projection}, and the use of higher-order quadrilateral FEs (8-node, 9-node FEs)~\cite {sarkar2025topology,diaz1995checkerboard,jog1996stability}, to name a few, have been used to circumvent such anomalies.

Topologically optimizing a CM results in a design in which the material is optimally distributed to perform a specific task under provided loading conditions. A design-dependent loading condition in TO  changes its direction, location, and/or magnitude as the optimization progresses, and thus poses several challenges in the TO framework~\cite{kumar2020topology,kumar2023topress}. Kumar et al.~\cite{kumar2020topology} propose a new approach to overcome those challenges, taking Darcy's law and the drainage term into consideration, and optimize 2D~\cite{kumar2020topology,kumar2024sorotop} and 3D CMs~\cite{kumar2021topology3DPaCM}.  Chen et al.~\cite{chen2018topology} synthesize a soft-robotic gripper under an actuation force, surface friction, and pressure loading conditions. de Souza et al.~\cite{de2020topology} address design-dependent TO problems for CMs applied to bending, linear, and inverter actuators. Lu and Tong~\cite{lu2021topology} present an extended algorithm in the TO framework for CMs under design-dependent pressure loads, wherein inverter and gripper CMs are designed. Pinskier et al.~\cite{pinskier2024diversity} provide a diversity-based TO approach for designing 3D multi-material grippers, wherein the design-dependent nature of the load is modelled using the Darcy method proposed in~\cite{kumar2020topology}. The Darcy method is used in~\cite{kumar2022topologyMM} and in~\cite{kumar2023topology} for designing multimaterial structures and CMs, respectively. In addition, MATLAB codes for 2D and 3D designs using the Darcy law are made publicly available for various problems settings by Kumar in~\cite{kumar2023topress,kumar2024sorotop,kumar2025topress3d}. Therefore, we adopt the Darcy method in this study for modelling the design-dependent nature of the pressure loads.

One very common problem in the optimized topology of a CM is the appearance of one-node-connected hinges. Along with this, imprecisions are observed in manufacturing the optimized design in macro, micro, and nano scales (milling, etching, e-beam lithography processes)~\cite{sigmund2009manufacturing}. A robust formulation proposed in~\cite{sigmund2009manufacturing,wang2011projection} seeks to mitigate the effects of these manufacturing tolerances. The worst-case scenario resulting from over-etching is referred to as the eroded design, whereas the intended configuration is termed the blueprint design.  Kumar et al.~\cite{kumar2021topology} design CMs for straining biological tissues using the robust formulation. Kumar and Langelaar~\cite{kumar2022topological} present the robust density-based TO framework for designing pressure-actuated compliant mechanisms (Pa-CMs). Kumar~\cite{kumar2024sorotop} introduces \texttt{SoRoTop} TO MATLAB code for optimizing pneumatic soft actuators and Pa-CMs, incorporating a robust formulation to account for the design-dependent nature of loads.

The remaining sections are arranged as follows. Sec. \ref{sec2} describes the quadrilateral finite elements used, Sec.~\ref{sec3} provides pressure load modeling using Darcy law, Sec. \ref{sec4} formulates the robust TO design, Sec. \ref{sec5} presents the results and discussions, and Sec. \ref{sec6} provides the concluding remarks.

\section{Quadrilateral Finite Elements}\label{sec2}
The Finite Element Method (FEM) is a widely used numerical approach for solving boundary value problems in TO. Among the quadrilateral elements (quad elements), four (Q4), eight (Q8), and nine-noded (Q9) elements are considered in this paper for discretizing the design domains. In general, using more nodes per mesh element improves the accuracy of the results. We compare the optimized results using the different elements mentioned herein.
\begin{figure}[H]
	\centering
	\begin{subfigure}[b]{0.30\textwidth}
		\centering
		\begin{tikzpicture}
            \fill [gray!30] (-1,-1) rectangle (1,1);
			\fill [black] (-1.05,-1.05) rectangle (1.05,-0.95);
			\fill [black] (1.05,-0.95) rectangle (0.95,1.05);
			\fill [black] (0.95,1.05) rectangle (-1.05,0.95);
			\fill [black] (-1.05,0.95) rectangle (-0.95,-1.05);
			\filldraw[blue] (-1,-1) circle (2.5pt) node[anchor=north east]{\large 1};
			\filldraw[blue] (1,-1) circle (2.5pt) node[anchor=north west]{\large 2};
			\filldraw[blue] (1,1) circle (2.5pt) node[anchor=south west]{\large 3};
			\filldraw[blue] (-1,1) circle (2.5pt) node[anchor=south east]{\large 4};
			\draw[brown, thick, -Stealth] (0,0) -- (2,0) node[anchor=south]{$\xi$};
			\draw[brown, thick, -Stealth] (0,0) -- (0,2) node[anchor=west]{$\eta$};
		\end{tikzpicture}
		\caption{}
		\label{fig:Q4}
	\end{subfigure}
	\begin{subfigure}[b]{0.30\textwidth}
		\centering
		\begin{tikzpicture}
            \fill [gray!30] (-1,-1) rectangle (1,1);
			\fill [black] (-1.05,-1.05) rectangle (1.05,-0.95);
			\fill [black] (1.05,-0.95) rectangle (0.95,1.05);
			\fill [black] (0.95,1.05) rectangle (-1.05,0.95);
			\fill [black] (-1.05,0.95) rectangle (-0.95,-1.05);
			\filldraw[blue] (-1,-1) circle (2.5pt) node[anchor=north east]{\large 1};
			\filldraw[blue] (1,-1) circle (2.5pt) node[anchor=north west]{\large 2};
			\filldraw[blue] (1,1) circle (2.5pt) node[anchor=south west]{\large 3};
			\filldraw[blue] (-1,1) circle (2.5pt) node[anchor=south east]{\large 4};
			\filldraw[blue] (0,-1) circle (2.5pt) node[anchor=north]{\large 5};
			\filldraw[blue] (1,0) circle (2.5pt) node[anchor=north west]{\large 6};
			\filldraw[blue] (0,1) circle (2.5pt) node[anchor=south east]{\large 7};
			\filldraw[blue] (-1,0) circle (2.5pt) node[anchor=east]{\large 8};
			\draw[brown, thick, -Stealth] (0,0) -- (2,0) node[anchor=south]{$\xi$};
			\draw[brown, thick, -Stealth] (0,0) -- (0,2) node[anchor=west]{$\eta$};
		\end{tikzpicture}
		\caption{}
		\label{fig:Q8}
	\end{subfigure}
	\begin{subfigure}[b]{0.30\textwidth}
		\centering
		\begin{tikzpicture}
            \fill [gray!30] (-1,-1) rectangle (1,1);
			\fill [black] (-1.05,-1.05) rectangle (1.05,-0.95);
			\fill [black] (1.05,-0.95) rectangle (0.95,1.05);
			\fill [black] (0.95,1.05) rectangle (-1.05,0.95);
			\fill [black] (-1.05,0.95) rectangle (-0.95,-1.05);
			\filldraw[blue] (-1,-1) circle (2.5pt) node[anchor=north east]{\large 1};
			\filldraw[blue] (1,-1) circle (2.5pt) node[anchor=north west]{\large 2};
			\filldraw[blue] (1,1) circle (2.5pt) node[anchor=south west]{\large 3};
			\filldraw[blue] (-1,1) circle (2.5pt) node[anchor=south east]{\large 4};
			\filldraw[blue] (0,-1) circle (2.5pt) node[anchor=north]{\large 5};
			\filldraw[blue] (1,0) circle (2.5pt) node[anchor=north west]{\large 6};
			\filldraw[blue] (0,1) circle (2.5pt) node[anchor=south east]{\large 7};
			\filldraw[blue] (-1,0) circle (2.5pt) node[anchor=east]{\large 8};
			\filldraw[blue] (0,0) circle (2.5pt) node[anchor= east]{\large 9};
			\draw[brown, thick, -Stealth] (0,0) -- (2,0) node[anchor=south]{$\xi$};
			\draw[brown, thick, -Stealth] (0,0) -- (0,2) node[anchor=west]{$\eta$};
		\end{tikzpicture}
		\caption{}
		\label{fig:Q9}
	\end{subfigure}
	\caption{Quadrilateral meshing elements : (a) Q4, (b) Q8, and (c) Q9. Numbers $1, 2, ..., 9$ indicate node numbers and $\xi$-$\eta$ is the natural coordinate system.}
	\label{fig: Different meshing elements}
\end{figure}
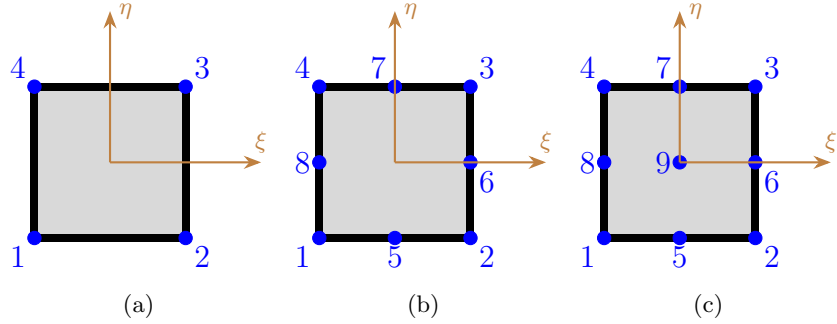
Figures \ref{fig:Q4}, \ref{fig:Q8}, and \ref{fig:Q9} represent the Q4, Q8, and Q9  elements~\cite{sarkar2025topology}. Blue dots represent the nodes, and the numbers in the same colour represent their number. $\xi \text{-} \eta$ is the natural coordinate system. The shape functions are evaluated according to the node numbers and their positions. Sarkar and Kumar~\cite{sarkar2025topology} describe these shape functions and their use.

\section{Pressure load modelling}\label{sec3}
Here, we describe the Darcy law with an additional drainage term in brief for the sake of completeness; the reader can refer to~\cite{kumar2020topology,kumar2023topress} for more details. The  flux, $q$ for the Darcy law can be determined as~\cite{kumar2020topology,kumar2023topress}
\begin{equation}
    q=-\frac{k}{\mu}\nabla p,
    \label{Darcy}
\end{equation}
where $\nabla p$ represents the pressure gradient, $k$ is the permeability of the medium, and $\mu$ is the viscosity of the fluid. The term $\frac{k}{\mu}$ is called flow coefficient $K$ and is defined as~\cite{kumar2020topology,kumar2023topress} 
\begin{equation}
    K(\bar{x}_i)=K_v \left(1-(1-\epsilon)H(\bar{x}_i,\beta_k,\eta_k)\right),
\end{equation}
where flow contrast is $\epsilon=\frac{K_s}{K_v}$. $K_v$ and $K_s$ are the flow coefficients of void and solid elements, respectively. $\bar{x}_i$ denotes the physical design variable for element~$i$. $\eta_k$ and $\beta_k$ are tuning parameters that determine the step location and slope gradient, respectively. $H(\bar{x}_i,\beta_k,\eta_k)$ is the Heaviside smooth function~\cite{kumar2020topology,kumar2023topress}.

To incorporate the pressure gradient at the solid-void interface, a drainage term $Q_\text{drain}$ is considered where, $Q_\text{drain} = -D(\bar{x}_i)(p - p_{ext})$. $D(\bar{x}_i)$, $p$, and $p_{ext}$ are the drainage coefficient, continuous pressure field, and external pressure, respectively~\cite{kumar2020topology,kumar2023topress}. Combining the drainage term with Darcy's law, the equilibrium equation is expressed as~\cite{kumar2020topology,kumar2023topress} 
\begin{equation}
    \nabla \cdot q - Q_\text{drain}=0.
    \label{balance1}
\end{equation}
In finite element modelling, Eq.~\ref{balance1} can be written as $\mathbf{A} \mathbf{P} = \mathbf{0}$, wherein $\mathbf{A}$ and $\mathbf{P}$ are the global flow matrix and global pressure field, respectively. This pressure field gives nodal forces, which in turn cause the deformation. The corresponding equation can be written as~\cite{kumar2020topology,kumar2023topress} 
\begin{equation}
    \mathbf{F} = \mathbf{K} \mathbf{U} = -\mathbf{TP}.
\end{equation}
$\mathbf{T}$, $\mathbf{F}$, $\mathbf{K}$, and $\mathbf{U}$ are the global transformation matrix, global force vector, global stiffness matrix, and global displacement vector, respectively.

\section{Topology optimization for robust design}\label{sec4}
The density-based TO approach is adopted, in which each finite element is assigned a design variable $x$. The material elastic stiffness is interpolated using the modified SIMP as
\begin{equation}
	E_i = E_v+(\bar{x}_i)^p(E_s-E_v),
    \label{Material_interpolation}
\end{equation}
where $E_s$ and $E_v$ are the elastic stiffness of an element with solid and void phases, respectively. $E_v$ is set to a very small percentage of $E_s$ ($E_v=10^{-6}E_s$), to avoid numerical instability. $E_i$ provides the interpolated material stiffness of element $i$, and $p$ is the SIMP penalization factor set to 3. $\bar{x}_i$, the projected variable obtained using the Heaviside projection filter, denotes the physical design variable for element $i$.
$\tilde{x}_i$ is the filtered design variable of element $i$, which is evaluated by the weighted average of the design variables inside a filter radius~\cite{bruns2001topology}.

Herein, we solve two of the most common CM design problems in TO: the inverter and gripper problems. The robust TO formulation presented in~\cite{kumar2024sorotop} is employed to achieve the optimized Pa-CMs. The optimization problem solved is as follows~\cite{kumar2024sorotop}

\begin{equation}
	\begin{rcases}
		\begin{aligned}
			&{\min_{\bm{\bar{x}}(\bm{\tilde{x}(\bm{x})})}} :\quad f_0 =  \max(\mathbf{L}^T\mathbf{U}_b,\,\mathbf{L}^T\mathbf{U}_e)\\
			&\mathrm{subjected to:}\\
            &\bm{\lambda}_{1r}:\,\,\mathbf{A}_r \mathbf{P}_r=\mathbf{0} \quad \quad \quad \quad \quad \quad(r=b,e)\\
            &\bm{\lambda}_{2r}:\,\,\mathbf{K}_r \mathbf{U}_r = \mathbf{F}_r = -\mathbf{T}\mathbf{P}_r \\
			&\Lambda_{b}:\,\,g_1\equiv\frac{V_b}{V^*}-1 \le 0 \\
            &\Lambda_{e}:\,\,g_2\equiv\frac{S_e}{S^*}-1 \le 0 \\
			&\quad\,\,\,\, 0 \leq x_{jr},\tilde{x}_{jr},\bar{x}_{jr} \leq 1 \quad(j=1,\,2,...,\,n)
		\end{aligned}
	\end{rcases}.
    \label{Optimization_eqn}
\end{equation}
$f_0$ is the objective function, which is to be minimized here. Subscripts $b$, $e$, and $ r$ represent blueprint, eroded design, and robust design, respectively. $\mathbf{L}$ is a virtual load vector whose all elements are zero except at the degree of freedom (DoF) of the desired output displacement, where the value is $1$. Therefore, $\mathbf{L}^T\mathbf{U}_b$ and $\mathbf{L}^T\mathbf{U}_e$ give the output displacements of blueprint and the eroded designs, respectively.

The robust TO design solves a min-max problem for CMs. We aim to minimize the function that gets the maximum value of these two output displacements. The volume constraint on the blueprint design is represented by $g_1$, where $V_b$ and $V^*$ denote the current volume and the total allowable volume of material for the CM, respectively. A spring with stiffness $k_s=1$ at the output DoF and a strain energy constraint $g_2$ applied to the eroded design for this robust formulation. The eroded design's current and total strain energy are indicated by $S_e$ and $S^*$, respectively. $\bm{\lambda}_{1r}$ (vector), $\bm{\lambda}_{2r}$ (vector), $\Lambda_{b}$ (scalar), and $\Lambda_{e}$ (scalar) are the Lagrangian multipliers associated with the corresponding constraints. The method of moving asymptotes (MMA) \cite{svanberg1987method} is employed to update the design variables after each iteration. We evaluate the sensitivities of the objective function and the constraints using the adjoint-variable method. The details of the sensitivity analyses can be found in~\cite{kumar2024sorotop}.

\section{Results and Discussions}\label{sec5}
\begin{figure}
	\begin{subfigure}{0.35\textwidth}
		\begin{tikzpicture}[scale=0.6]
			\makeatletter
			\pgfdeclarepatternformonly[\LineSpace,\LineThickness]{custom hatch}
			{\pgfqpoint{0pt}{0pt}}{\pgfqpoint{\LineSpace}{\LineSpace}}
			{\pgfqpoint{\LineSpace}{\LineSpace}}
			{
				\pgfsetlinewidth{\LineThickness}
				\pgfpathmoveto{\pgfqpoint{0pt}{0pt}}
				\pgfpathlineto{\pgfqpoint{\LineSpace}{\LineSpace}}
				\pgfusepath{stroke}
			}
			\makeatother
			\tikzset{
				LineSpace/.store in=\LineSpace,
				LineSpace=2pt,
				LineThickness/.store in=\LineThickness,
				LineThickness=0.4pt
			}
			
			\fill[pattern=custom hatch] (0,-0.3) rectangle (0.8,0);
			\fill[pattern=custom hatch] (0,8) rectangle (0.8,8.3);
			
			\fill [gray!50] (0,0) rectangle (8,8);
            \fill [brown] (0,7.6) rectangle (0.8,8);
            \fill [brown] (0,0.4) rectangle (0.8,0);
            \fill [brown] (7.2,3.6) rectangle (8,4.4);
			\draw[black, very thick] (0,0) -- (8,0);
			\draw[black, very thick] (8,0) -- (8,8);
			\draw[black, very thick] (8,8) -- (0,8);
			\draw[black, very thick] (0,8) -- (0,0);
			\dimline[color=blue,extension start length=2pt, extension end length=2pt,label style={fill=none, yshift=-8pt}]{(0,-0.6)}{(8,-0.6)}{$L_x$};
            \dimline[color=blue,extension start length=5pt, extension end length=5pt,label style={fill=none, yshift=-8pt}]{(0,7.4)}{(0.8,7.4)}{$\frac{L_x}{20}$};
            \dimline[color=blue,extension start length=5pt, extension end length=5pt,label style={fill=none, yshift=8pt}]{(0,0.6)}{(0.8,0.6)}{\tiny$\frac{L_x}{20}$};
			\dimline[color=blue,extension start length=5pt, extension end length=5pt,label style={fill=none, yshift=-8pt}]{(1,7.6)}{(1,8)}{$\frac{L_y}{40}$};
			\dimline[color=blue,extension start length=5pt, extension end length=5pt,label style={fill=none, yshift=-8pt}]{(1,0)}{(1,0.4)}{$\frac{L_y}{40}$};
            \dimline[color=blue,extension start length=5pt, extension end length=5pt,label style={fill=none, yshift=8pt}]{(7.2,4.6)}{(8,4.6)}{$\frac{L_x}{20}$};
            \dimline[color=blue,extension start length=5pt, extension end length=5pt,label style={fill=none, yshift=8pt}]{(7,3.6)}{(7,4.4)}{\tiny$\frac{L_y}{20}$};
            
            \draw[red, very thick, -Stealth] (-1.5,8) -- (0,8);
            \draw[red, very thick, -Stealth] (-1.5,7) -- (0,7);
            \draw[red, very thick, -Stealth] (-1.5,6) -- (0,6);
            \draw[red, very thick, -Stealth] (-1.5,5) -- (0,5);
            \draw[red, very thick, -Stealth] (-1.5,4) -- (0,4);
            \draw[red, very thick, -Stealth] (-1.5,3) -- (0,3);
            \draw[red, very thick, -Stealth] (-1.5,2) -- (0,2);
            \draw[red, very thick, -Stealth] (-1.5,1) -- (0,1);
            \draw[red, very thick, -Stealth] (-1.5,0) -- (0,0);
			\dimline[color=blue,extension start length=5pt, extension end length=5pt,label style={fill=none, yshift=-6pt}]{(8.8,0)}{(8.8,4)}{$L_y/2$};
			\dimline[color=blue,extension start length=5pt, extension end length=5pt,label style={fill=none, yshift=-6pt}]{(8.8,4)}{(8.8,8)}{$L_y/2$};
			\draw[green, very thick, -Stealth] (9.8,4) -- (8,4);
            \draw[decorate, decoration={coil, aspect=0.3, segment length=1.5mm, amplitude=2mm}] (8,4) -- (9.8,4);
            \fill [black] (9.6,3.8) rectangle (9.8,4.2);
            \node at (9.6,4.8) {$k_{\mathrm{s}}$};
			\node at (-1,3.5) {$p$};
			\node at (8.5,3.3) {$d$};
		\end{tikzpicture}
		\caption{Inverter}
		\label{Inverter}
	\end{subfigure}
	\hspace{2.5cm}
	\begin{subfigure}{0.35\textwidth}
		\begin{tikzpicture}[scale=0.6]			
			\tikzset{
				LineSpace/.store in=\LineSpace,
				LineSpace=2pt,
				LineThickness/.store in=\LineThickness,
				LineThickness=0.4pt
			}
			
			\fill[pattern=custom hatch] (0,-0.3) rectangle (0.8,0);
			\fill[pattern=custom hatch] (0,8) rectangle (0.8,8.3);
			
            \fill [gray!50] (0,0) rectangle (8,8);
            \fill [brown] (0,7.6) rectangle (0.8,8);
            \fill [brown] (0,0.4) rectangle (0.8,0);
			\fill [white] (6.4,3) rectangle (8,5);
			\fill [brown] (6.4,3) rectangle (8,3.2);
			\fill [brown] (6.4,5) rectangle (8,4.8);
            \dimline[color=blue,extension start length=5pt, extension end length=5pt,label style={fill=none, yshift=-8pt}]{(0,7.4)}{(0.8,7.4)}{$\frac{L_x}{20}$};
            \dimline[color=blue,extension start length=5pt, extension end length=5pt,label style={fill=none, yshift=8pt}]{(0,0.6)}{(0.8,0.6)}{\tiny$\frac{L_x}{20}$};
			\dimline[color=blue,extension start length=5pt, extension end length=5pt,label style={fill=none, yshift=-8pt}]{(1,7.6)}{(1,8)}{$\frac{L_y}{40}$};
			\dimline[color=blue,extension start length=5pt, extension end length=5pt,label style={fill=none, yshift=-8pt}]{(1,0)}{(1,0.4)}{$\frac{L_y}{40}$};
			
			\draw[black, very thick] (0,0) -- (8,0);
			\draw[black, very thick] (8,0) -- (8,3.2);
			\draw[black, very thick] (8,4.8) -- (8,8);
			\draw[black, very thick] (8,8) -- (0,8);
			\draw[black, very thick] (0,8) -- (0,0);
			\draw[black, very thick] (6.4,3) -- (6.4,5);
			
			\draw[dash pattern=on 16pt off 2pt on 1pt off 2pt] (0,4) -- (9,4);
			\dimline[color=blue,extension start length=2pt, extension end length=2pt,label style={fill=none, yshift=-6pt}]{(0,-0.6)}{(8,-0.6)}{$L_x$};
			\dimline[color=blue,extension start length=5pt, extension end length=5pt,label style={fill=none, yshift=-6pt}]{(8.8,0)}{(8.8,4)}{$L_y/2$};
			\dimline[color=blue,extension start length=5pt, extension end length=5pt,label style={fill=none, yshift=-6pt}]{(8.8,4)}{(8.8,8)}{$L_y/2$};
			\dimline[color=blue,extension start length=5pt, extension end length=5pt,label style={fill=none, yshift=4pt}]{(7,0)}{(7,3)}{\tiny$0.375L_y$};
			\dimline[color=blue,extension start length=5pt, extension end length=5pt,label style={fill=none, yshift=4pt}]{(7,5)}{(7,8)}{\tiny$0.375L_y$};
			\dimline[color=blue,extension start length=5pt, extension end length=5pt,label style={fill=none, yshift=4pt}]{(7,3.2)}{(7,4.8)}{\tiny$0.2L_y$};
			
			\draw[red, very thick, -Stealth] (-1.5,8) -- (0,8);
            \draw[red, very thick, -Stealth] (-1.5,7) -- (0,7);
            \draw[red, very thick, -Stealth] (-1.5,6) -- (0,6);
            \draw[red, very thick, -Stealth] (-1.5,5) -- (0,5);
            \draw[red, very thick, -Stealth] (-1.5,4) -- (0,4);
            \draw[red, very thick, -Stealth] (-1.5,3) -- (0,3);
            \draw[red, very thick, -Stealth] (-1.5,2) -- (0,2);
            \draw[red, very thick, -Stealth] (-1.5,1) -- (0,1);
            \draw[red, very thick, -Stealth] (-1.5,0) -- (0,0);
			\draw[green, very thick, -{Stealth[scale=0.5]}] (8,3.2) -- (8,3.8);
			\draw[green, very thick, -{Stealth[scale=0.5]}] (8,4.8) -- (8,4.2);
            \draw[decorate, decoration={coil, aspect=0.3, segment length=1mm, amplitude=0.6mm}] (8,3.2) -- (8,3.8);
            \fill [black] (8.2,3.8) rectangle (7.8,3.9);
			\draw[green, very thick, -{Stealth[length=1.8mm, width=2mm]}] (8,4.8) -- (8,4.2);
            \draw[decorate, decoration={coil, aspect=0.3, segment length=1mm, amplitude=0.6mm}] (8,4.8) -- (8,4.2);
            \fill [black] (8.2,4.2) rectangle (7.8,4.1);
            \node at (8.4,3.2) {\small$d$};
            \node at (8.4,4.8) {\small$d$};
			\node at (7.5,3.6) {$k_{\mathrm{s}}$};
			\node at (7.5,4.4) {$k_{\mathrm{s}}$};
			\node at (-1,3.5) {$p$};
		\end{tikzpicture}
		\caption{Gripper}
		\label{Gripper}
	\end{subfigure}
	\caption{The design domains of (a) an inverter, and (b) a gripper}
	\label{Compliant_mechanism_2D}
\end{figure}
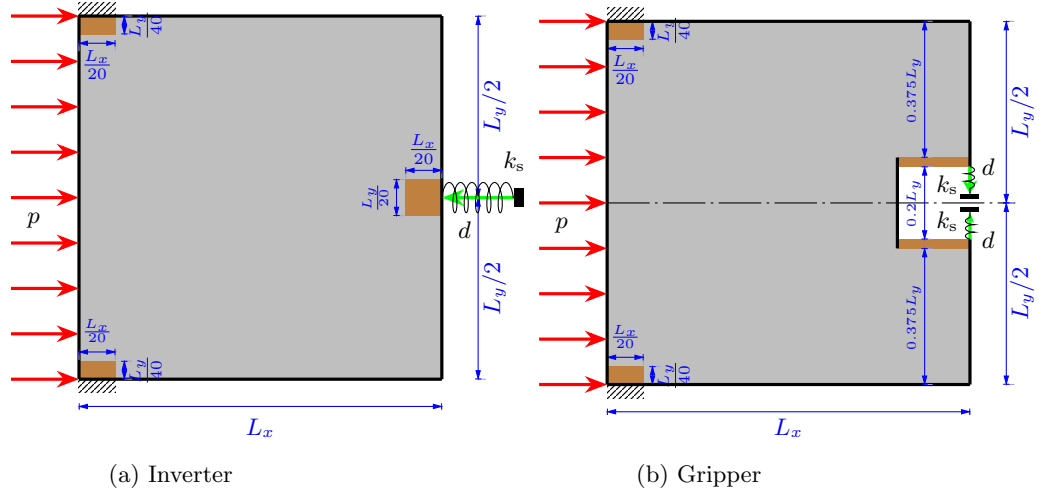

Figures \ref{Inverter} and \ref{Gripper} show the design domain of an inverter and of a gripper mechanism, respectively. A uniform pressure load (design-dependent load) $p$ acts upon the left edges of the CMs, as shown by arrows in red. Green arrows indicate the desired direction of the output displacement~$d$. $L_x$ and $L_y$ represent the design domain's length along the \(x\) direction and width along the \(y\) direction, respectively. The brown regions indicate the non-design solid portions of the CM.

The design domains of CMs are discretized by $200$ FEs along both the \(x\) and \(y\) directions. The symmetric top half portions are considered for optimization for both the inverter and gripper mechanisms. The maximum number of optimization iterations is fixed to 400. Young's modulus of the material is normalized to~1.

Two cases with different strain energy constraints ($S_e/S^*=0.95$ and $0.80$) are considered for optimizing the inverter mechanism. The permissible volume fraction of the blueprint is set as $V_b/V^*=0.25$ for both cases. The filter radius is taken as 6.4. For the robust formulation, $\Delta\eta = 0.15$ is taken. The projection filter parameter starts from 1 and is doubled every 50 iterations until it reaches 128. The flow parameters, $\eta_k$ is set to 0.1 and $\beta_k$ is taken as 8~\cite{kumar2024sorotop}.

For the gripper mechanism, $S_e/S^*$ is set to 0.90 for optimization purposes. The permissible volume fraction is taken as $0.30$ ($=V_b/V^*$) for the blueprint, and the filter radius is set to 5.6. For the robust formulation, $\Delta\eta = 0.10$ is taken~\cite{wang2011projection,kumar2024sorotop}. The flow parameter $\beta_k$ is taken as 10. Other parameters are the same as those used for the inverter mechanism. 

Table~\ref{Inverter_table_SE95}, Table~\ref{Inverter_table_SE80}, and Table~\ref{Gripper_table} show the optimized results for the compliant mechanisms under design-dependent pressure load. The first column indicates the quadrilateral elements (QE) used, i.e., Q4, Q8, or Q9. The second column represents the value of the objective function ($f_0$) for the blueprint and the eroded mechanisms. The third column depicts the volume fractions ($V/V^*$) of the material for the blueprint and the eroded mechanisms. The last two columns illustrate the optimized blueprint and eroded mechanisms. The red arrows indicate the pressure load acting perpendicular to the surface on the left side.

Table \ref{Inverter_table_SE95} shows the optimized results of the inverter for $S_e/S^*=0.95$. The value of the objective is better with Q8 for the blueprint mechanisms and with Q4 for the eroded mechanisms. This indicates more flexibility at the output DoF of the blueprint design with Q8 and the eroded design with Q4. The negative sign indicates that the displacement is initially in the negative direction of the applied pressure load. Therefore, the lower $f_0$ value represents a greater output displacement, better results, and increased flexibility.
A variation in thickness is noticeable in Table \ref{Inverter_table_SE95} for different quad elements. The final volume fraction achieved by the blueprint designs is 0.25, the maximum limit set for the optimization. The eroded designs obtain a volume fraction of 0.227 for Q4, Q8 and Q9.
 
\begin{table}[H] 
\centering
\caption{Optimized design for the inverter with pressure load ($S_e/S^*=0.95$)}
\label{Inverter_table_SE95}
\begin{tabular}{|C{1cm}|C{1.6cm}|C{1cm}|C{4cm}|C{4cm}|}
\hline
\textbf{QE} & $\bm{f_0}$ & $\bm{V/V^*}$ & \textbf{Blueprint} & \textbf{Eroded} \\ \hline
Q4 & \shortstack{$-115.4893$ \\ $-123.4772$} & \shortstack{$0.250$ \\ $0.227$} & \rule{0pt}{3.3cm} \includegraphics[scale=0.35]{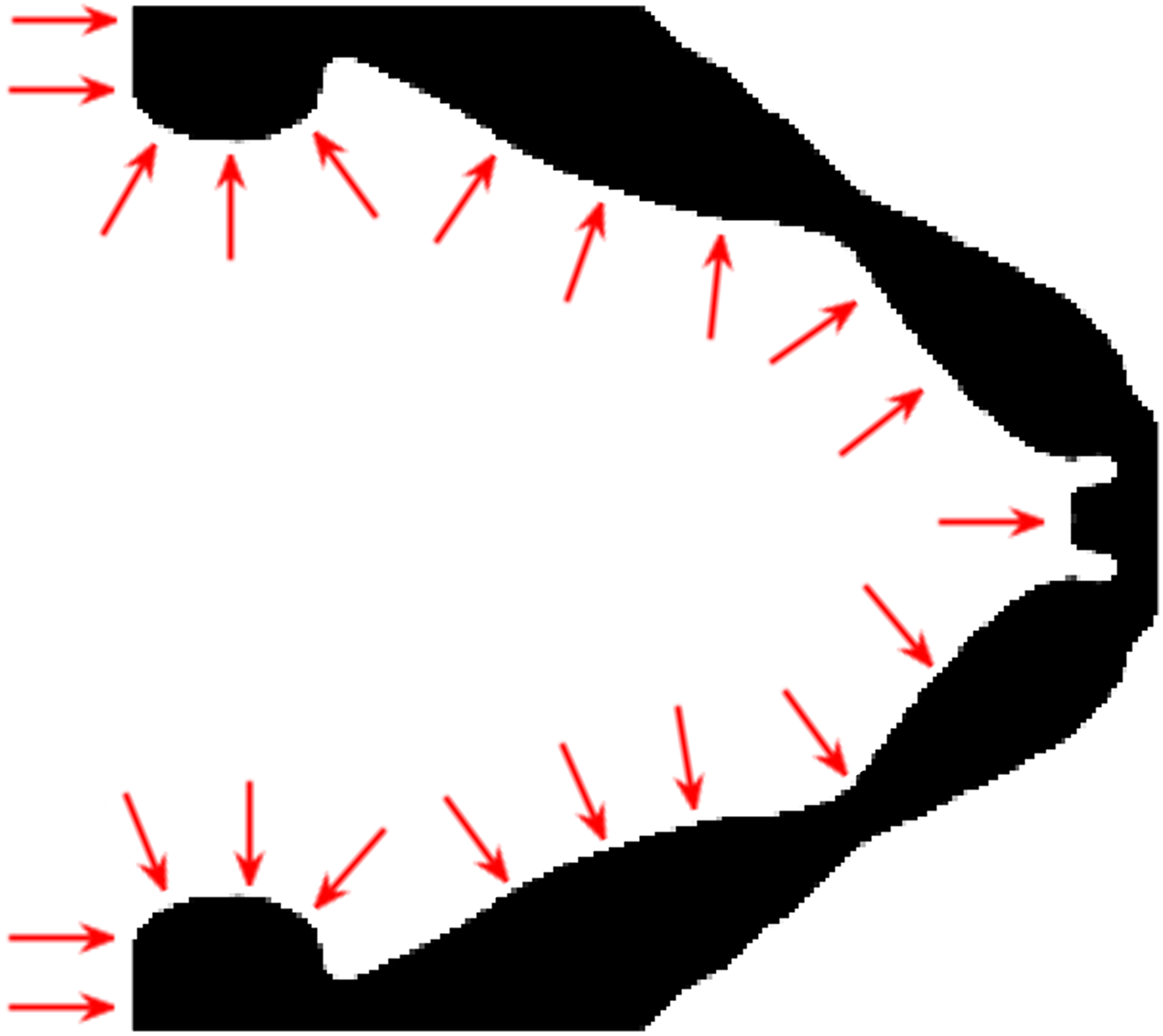} & \rule{0pt}{3.3cm} \includegraphics[scale=0.35]{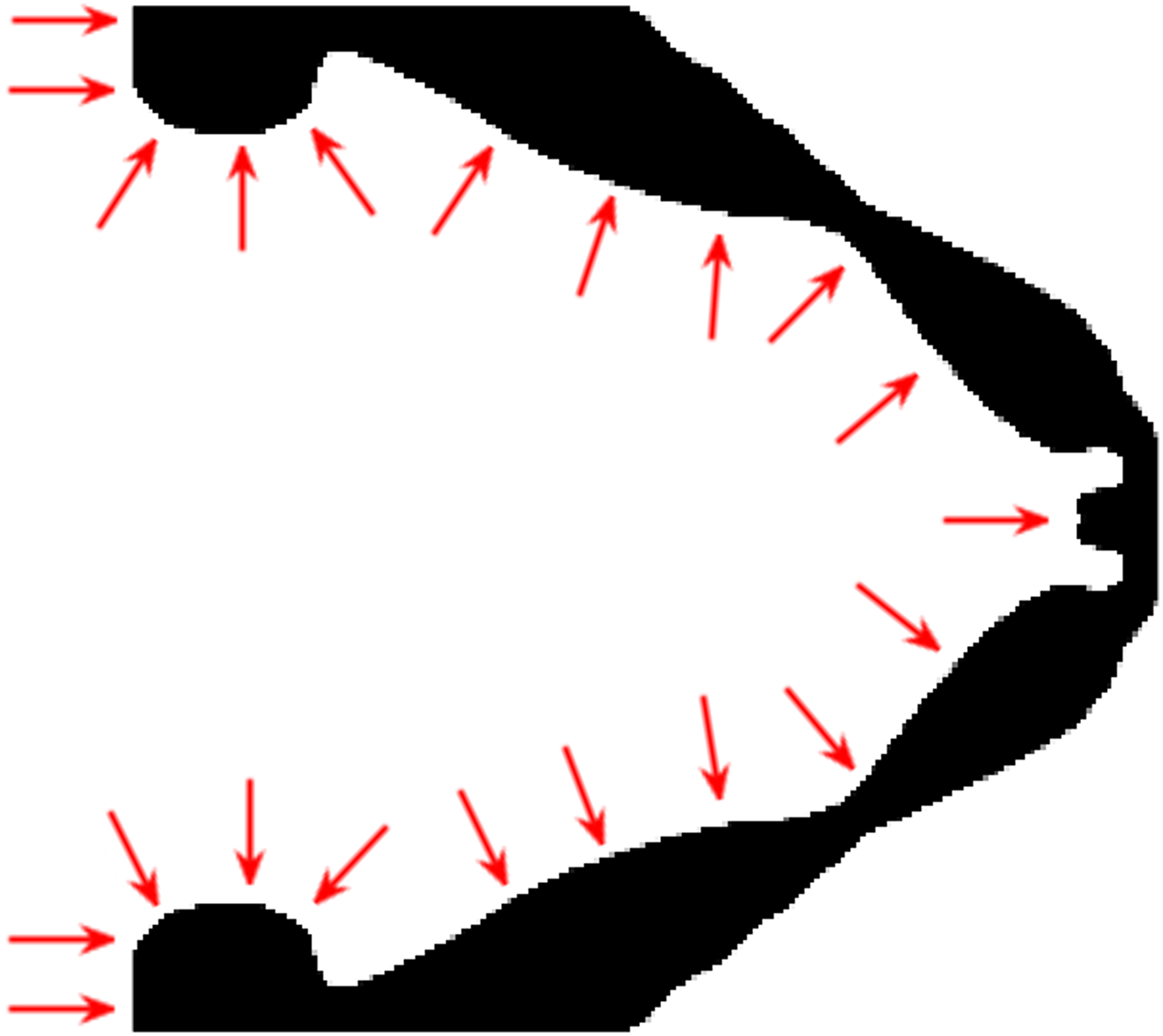} \\ \hline
Q8 & \shortstack{$-117.0731$ \\ $-120.1812$} & \shortstack{$0.250$ \\ $0.227$} & \rule{0pt}{3.3cm} \includegraphics[scale=0.35]{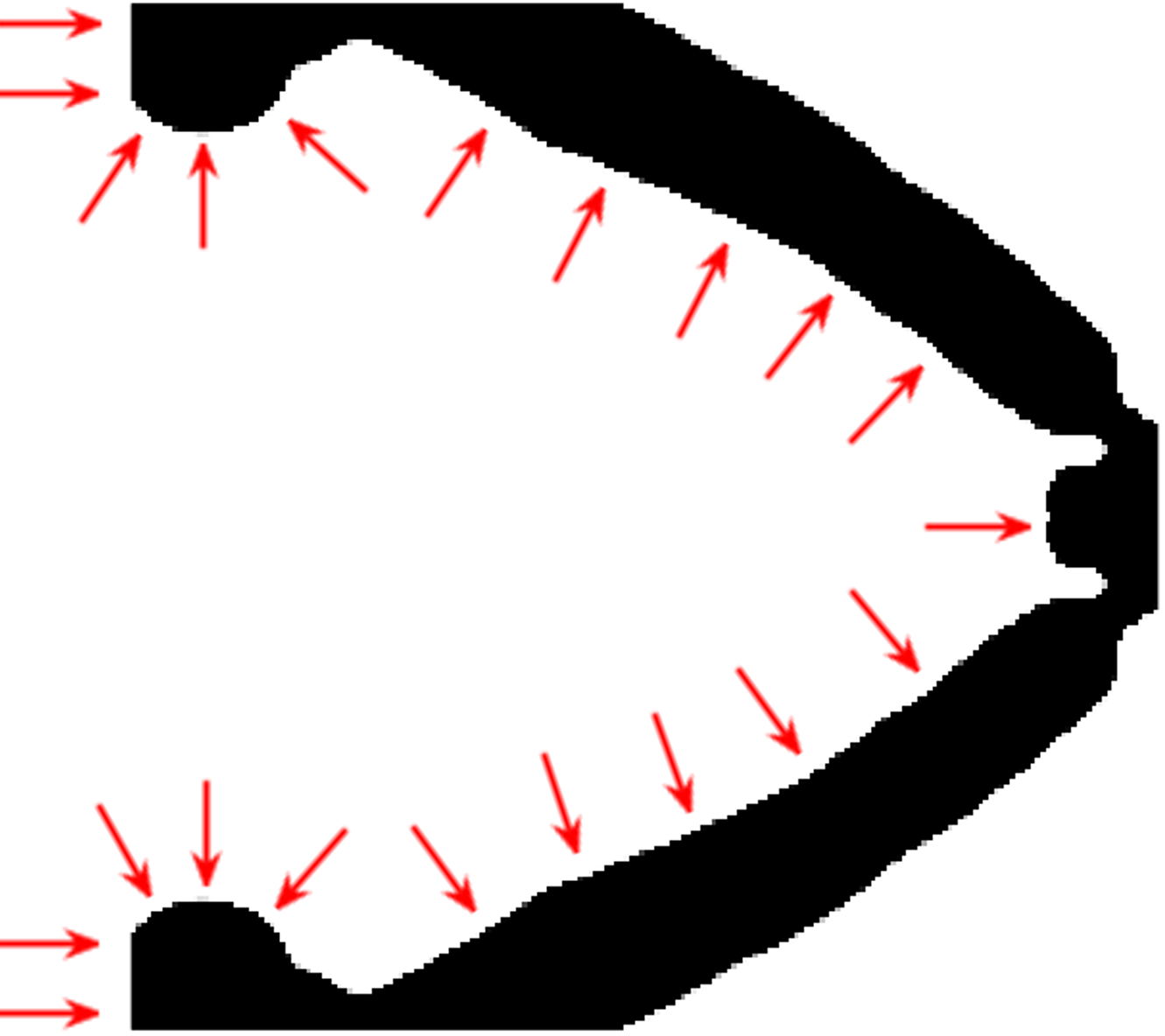} & \rule{0pt}{3.3cm} \includegraphics[scale=0.35]{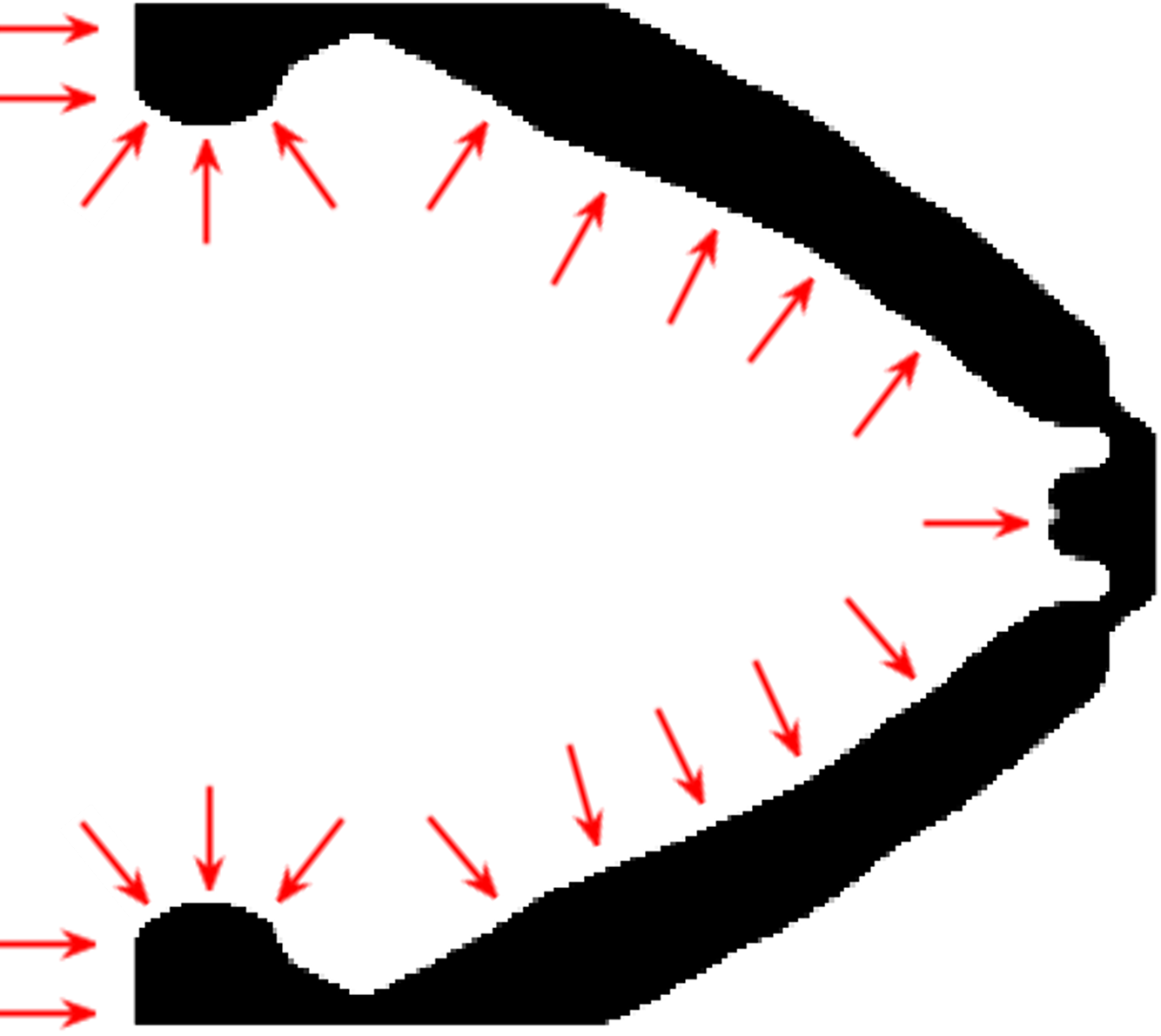} \\ \hline
Q9 & \shortstack{$-116.5421$ \\ $-118.6900$} & \shortstack{$0.250$ \\ $0.227$} & \rule{0pt}{3.3cm} \includegraphics[scale=0.35]{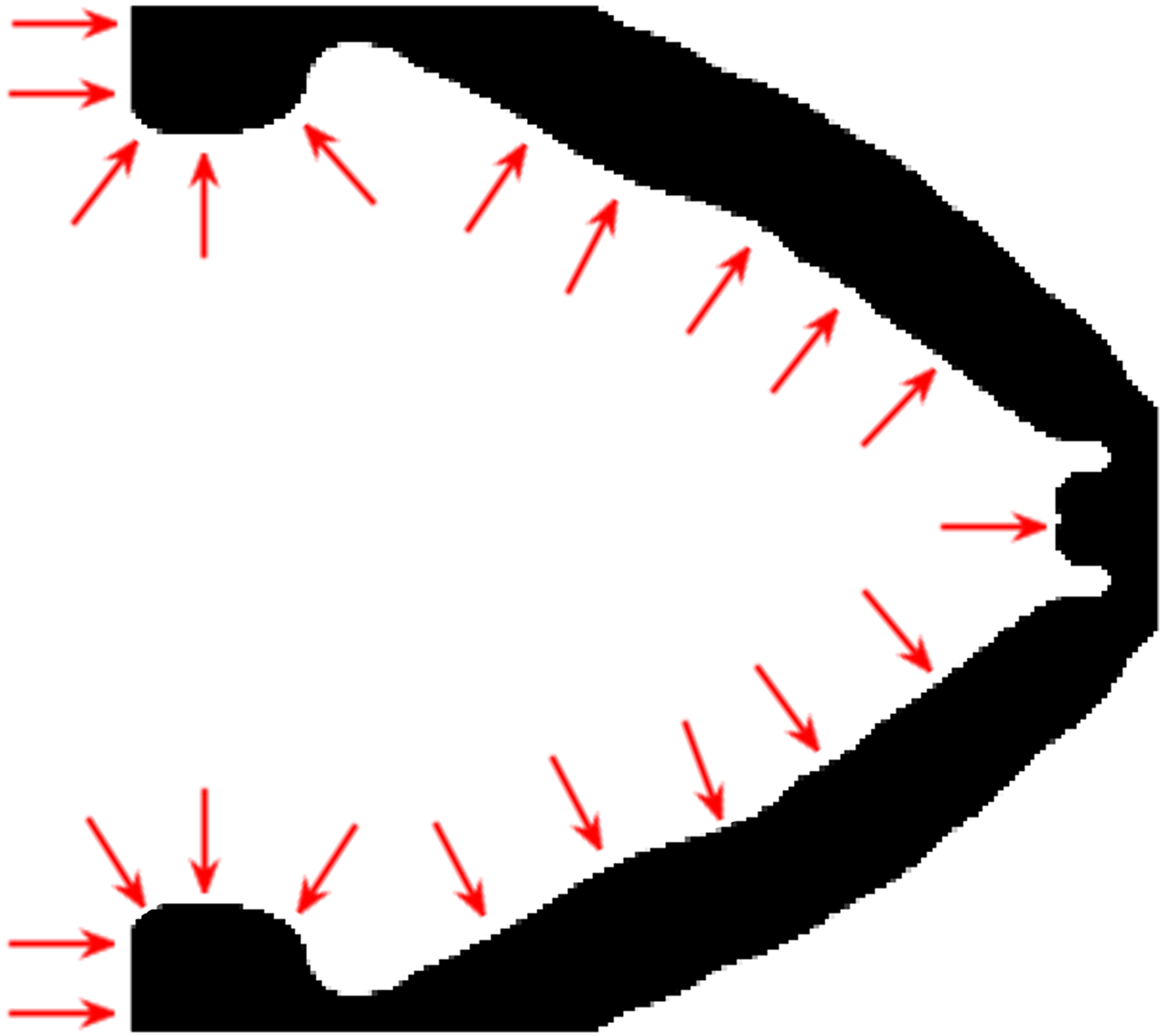} & \rule{0pt}{3.3cm} \includegraphics[scale=0.35]{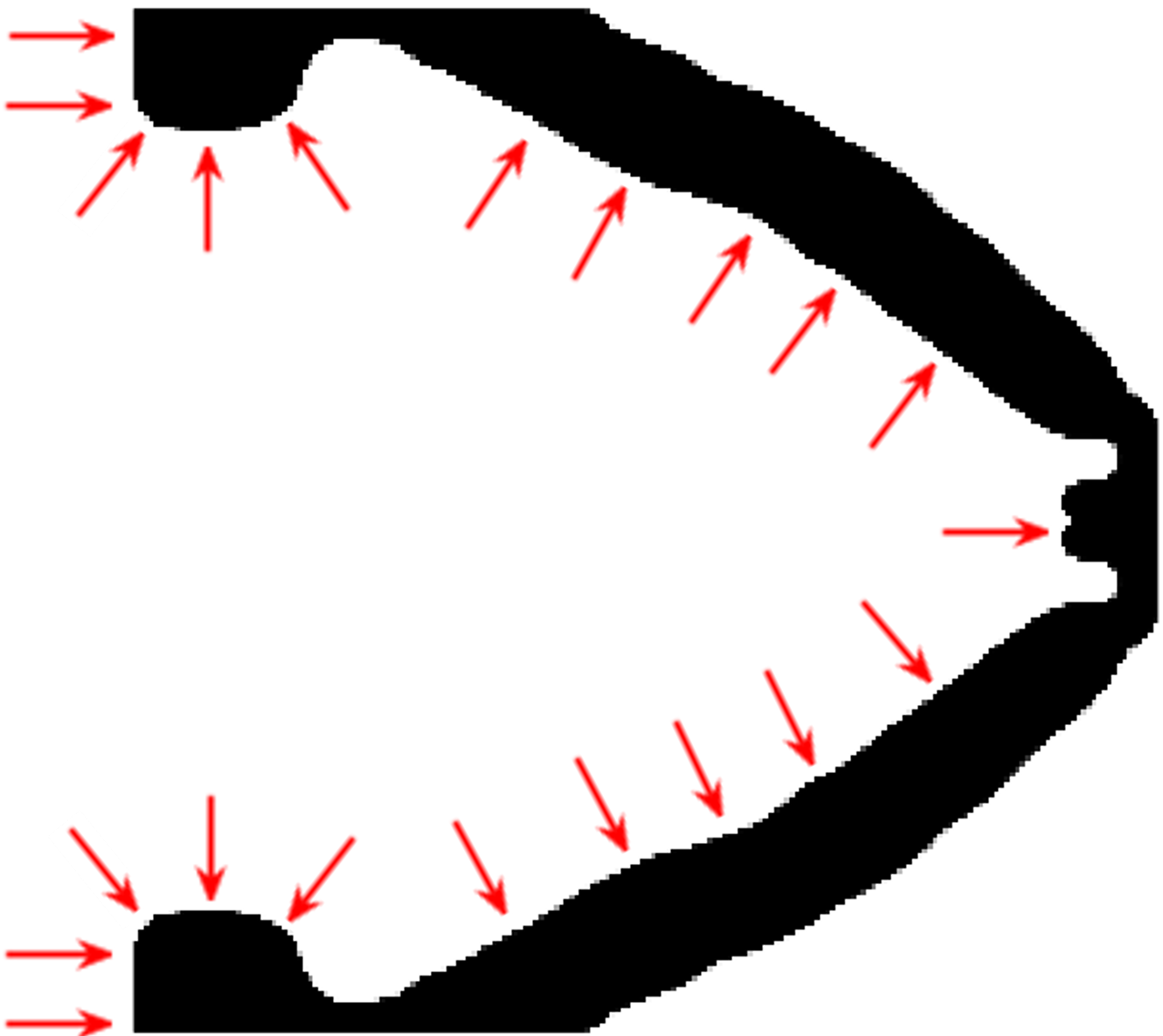} \\
\hline
\end{tabular}
\end{table}

Table \ref{Inverter_table_SE80} shows the optimized results of the inverter for $S_e/S^*=0.80$. Q8 has better objective value for blueprint and eroded mechanisms than Q4 and Q9, that is, optimized designs with Q8 are more flexible. The variations in thickness in the optimized results make the topology different for Q4, Q8, and Q9. The permissible volume limit is achieved by the blueprint designs in all of these quad element cases. In addition, as expected, the volume fractions achieved by eroded designs are less than those of the blueprint designs.

\begin{table}[] 
\centering
\caption{Optimized design for the inverter with pressure load ($S_e/S^*=0.80$)}
\label{Inverter_table_SE80}
\begin{tabular}{|C{1cm}|C{1.6cm}|C{1cm}|C{4cm}|C{4cm}|}
\hline
\textbf{QE} & $\bm{f_0}$ & $\bm{V/V^*}$ & \textbf{Blueprint} & \textbf{Eroded} \\ \hline
Q4 & \shortstack{$-101.9138$ \\ $-106.0701$} & \shortstack{$0.250$ \\ $0.226$} & \rule{0pt}{3.3cm} \includegraphics[scale=0.35]{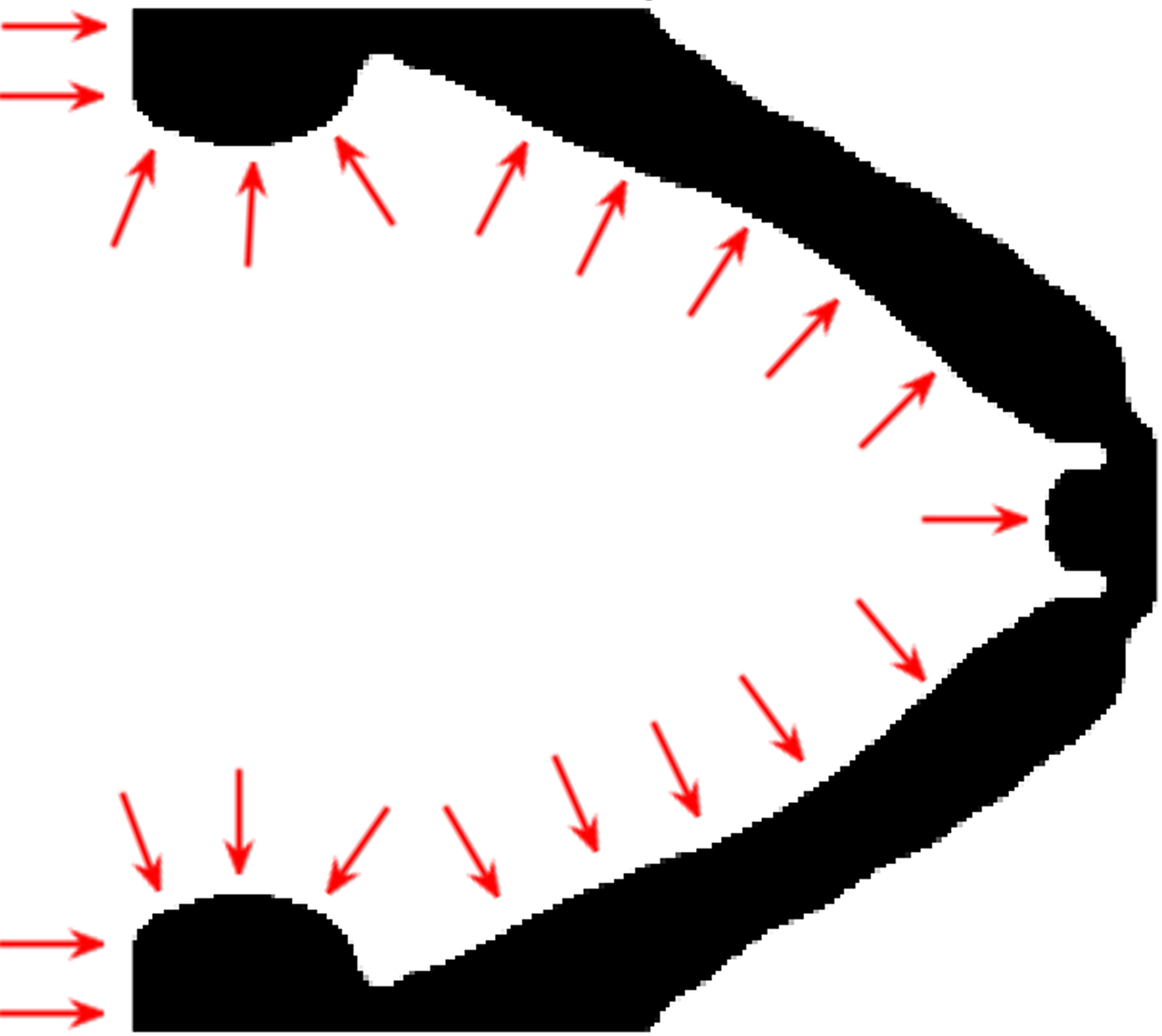} & \rule{0pt}{3.3cm} \includegraphics[scale=0.35]{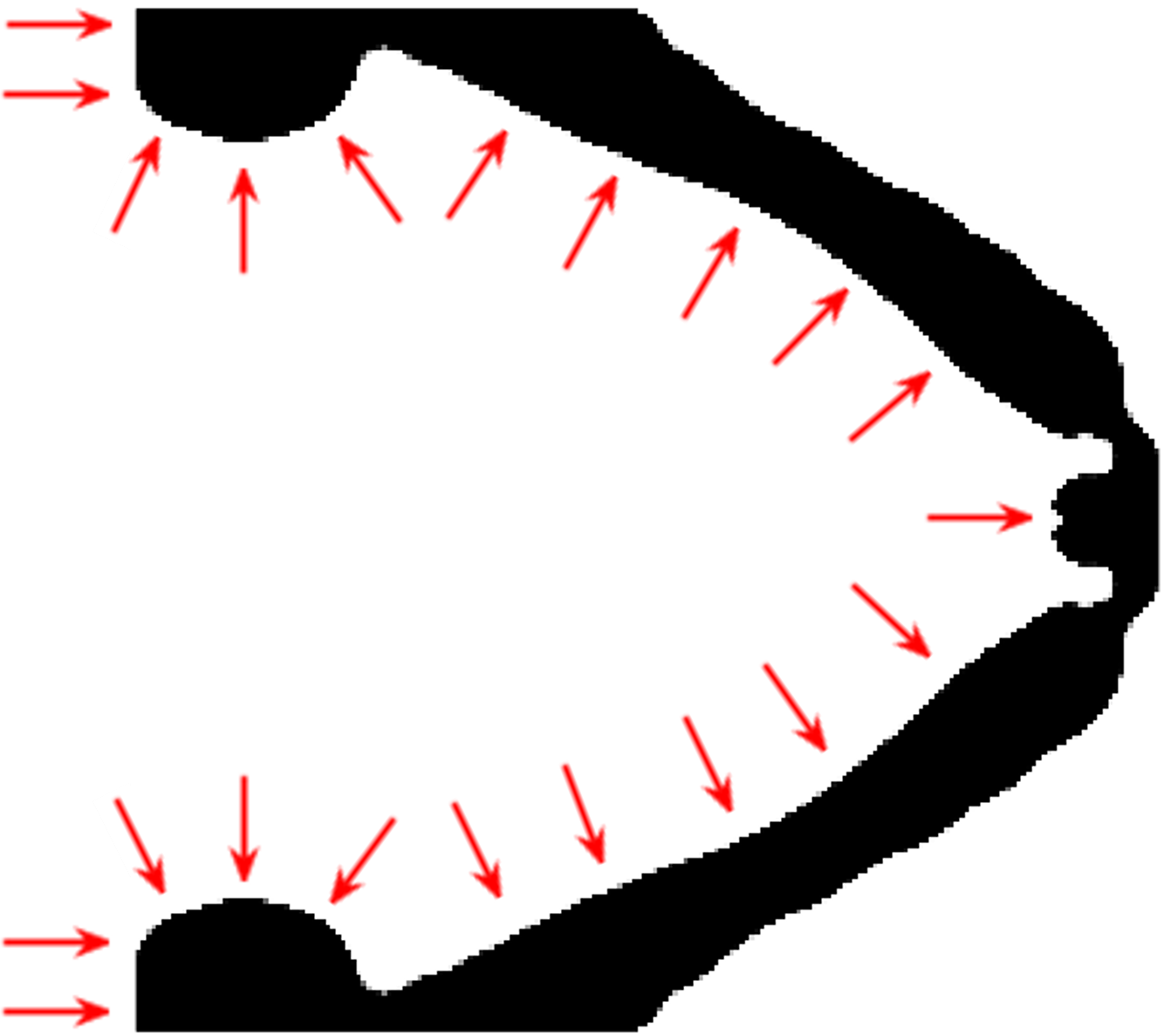} \\ \hline
Q8 & \shortstack{$-113.4375$ \\ $-117.3105$} & \shortstack{$0.250$ \\ $0.229$} & \rule{0pt}{3.3cm} \includegraphics[scale=0.35]{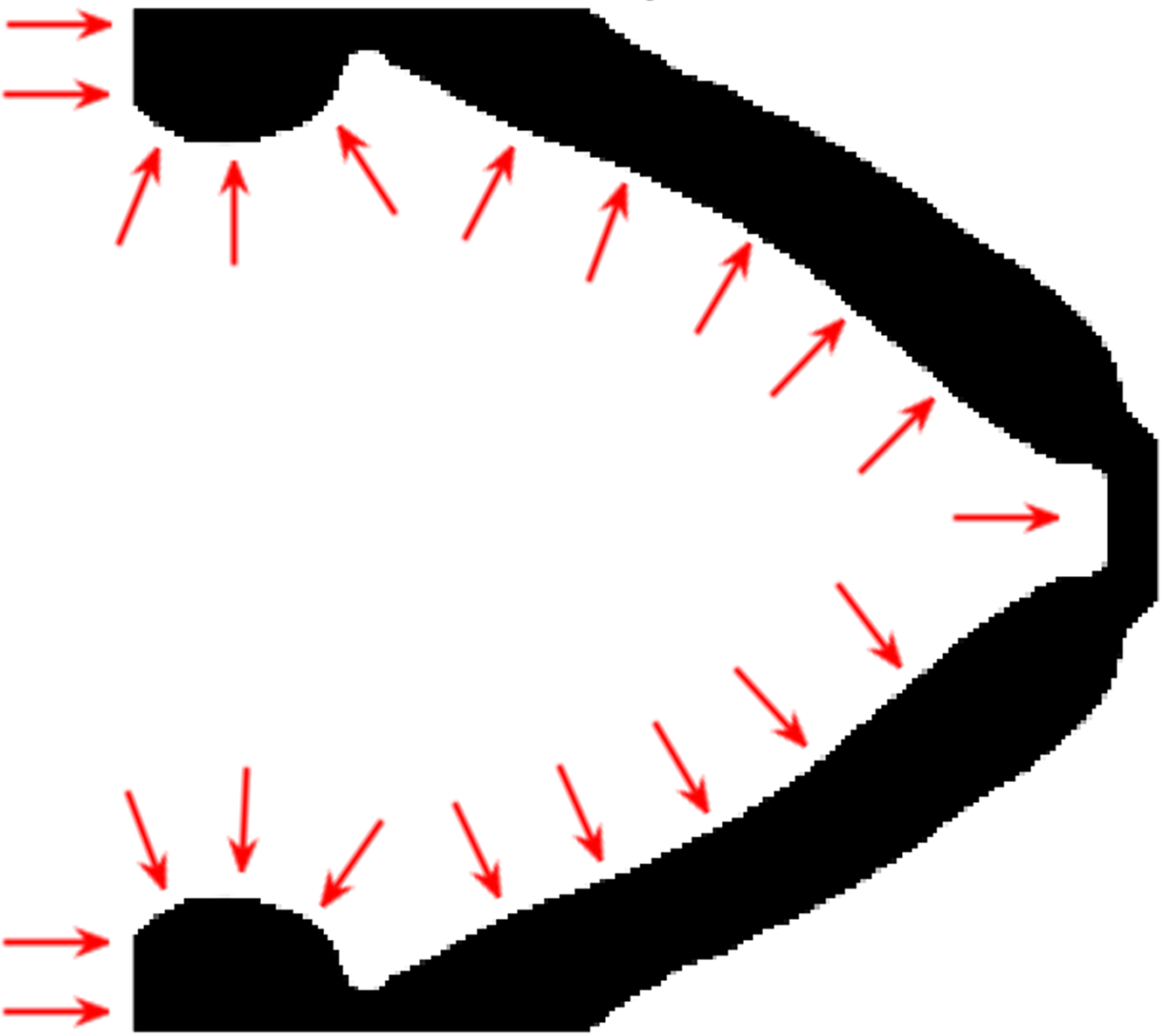} &
\rule{0pt}{3.3cm} \includegraphics[scale=0.35]{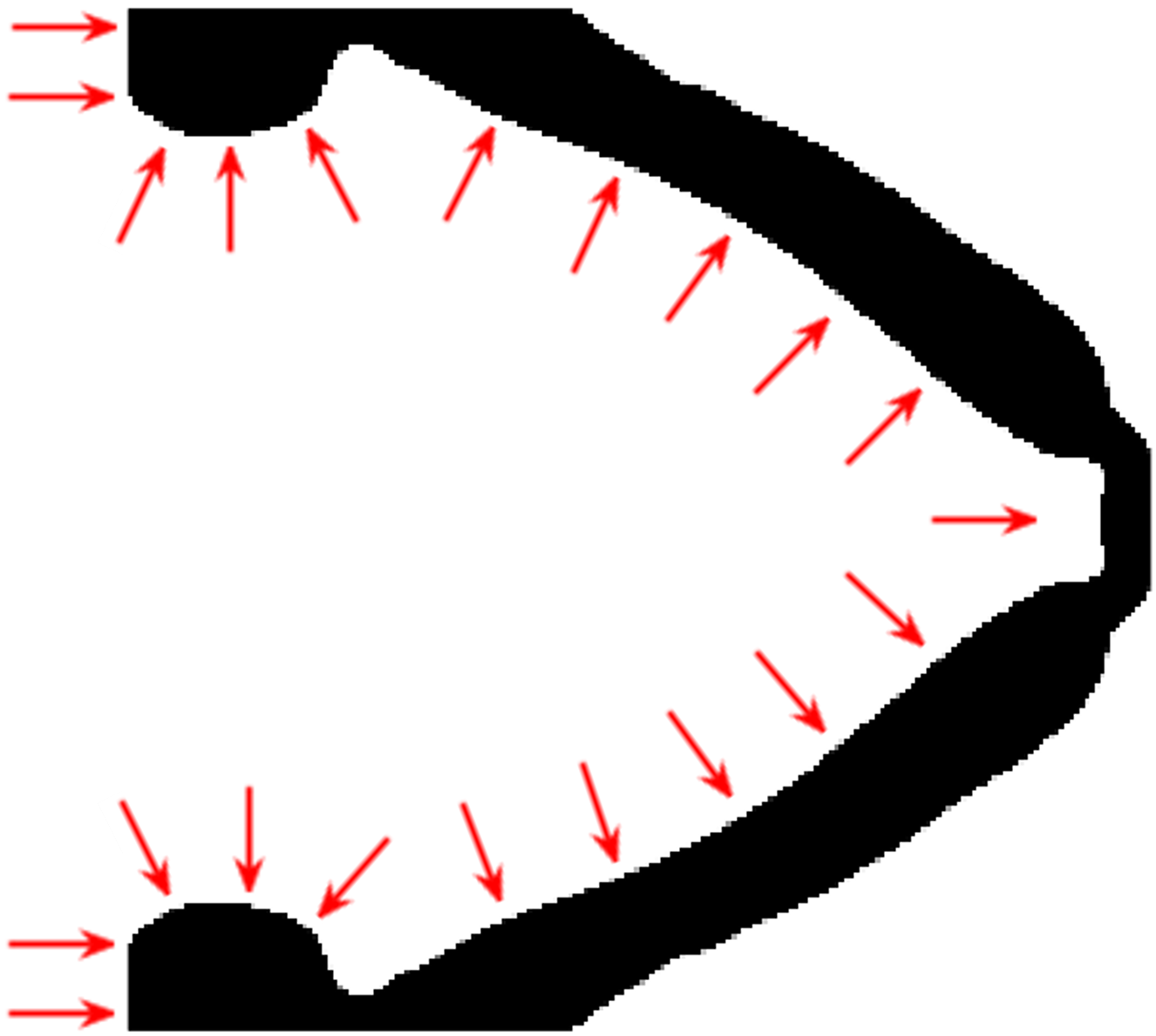} \\ \hline
Q9 & \shortstack{$-107.7355$ \\ $-111.6114$} & \shortstack{$0.250$ \\ $0.226$} & \rule{0pt}{3.3cm} \includegraphics[scale=0.35]{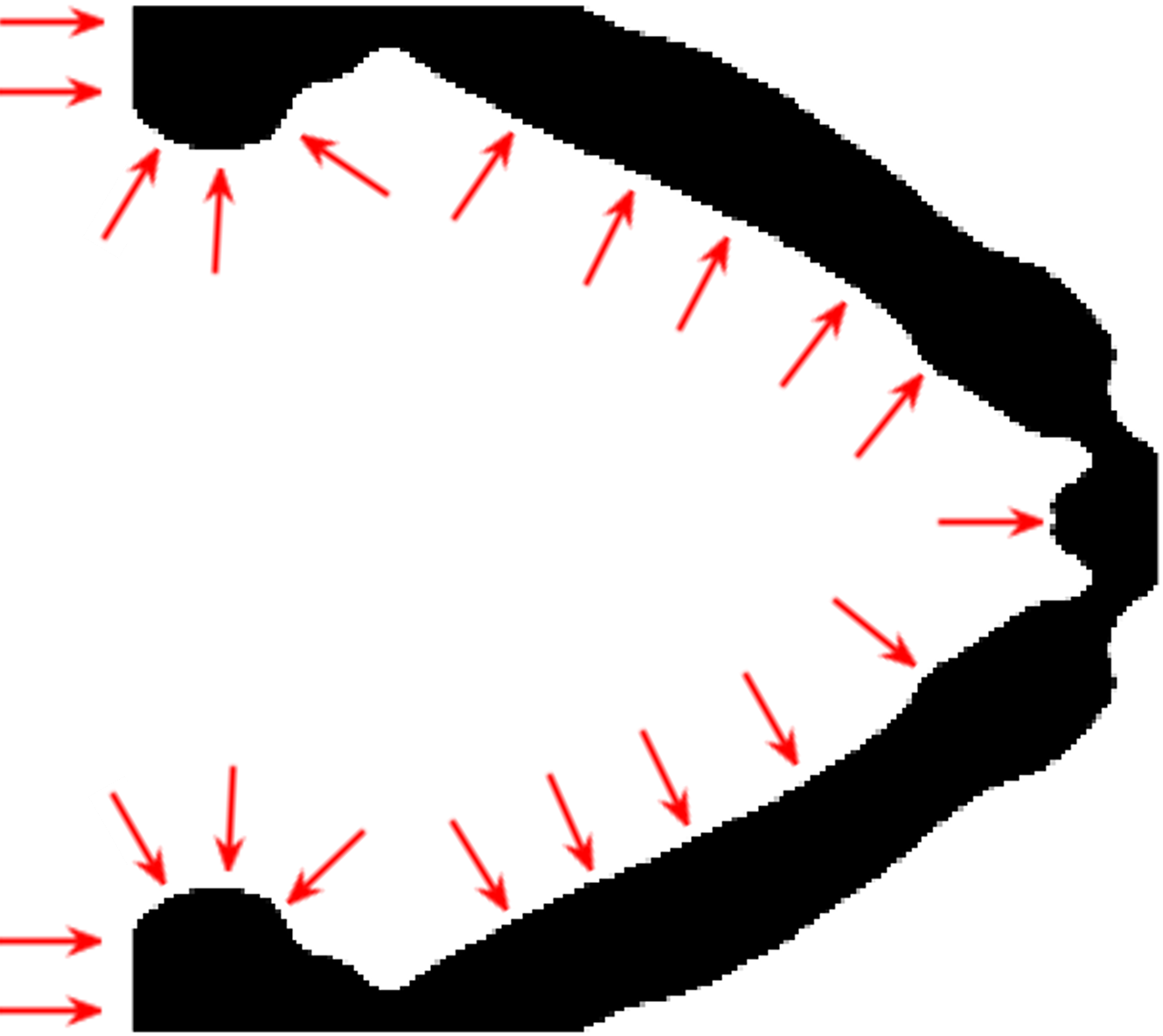} & \rule{0pt}{3.3cm} \includegraphics[scale=0.35]{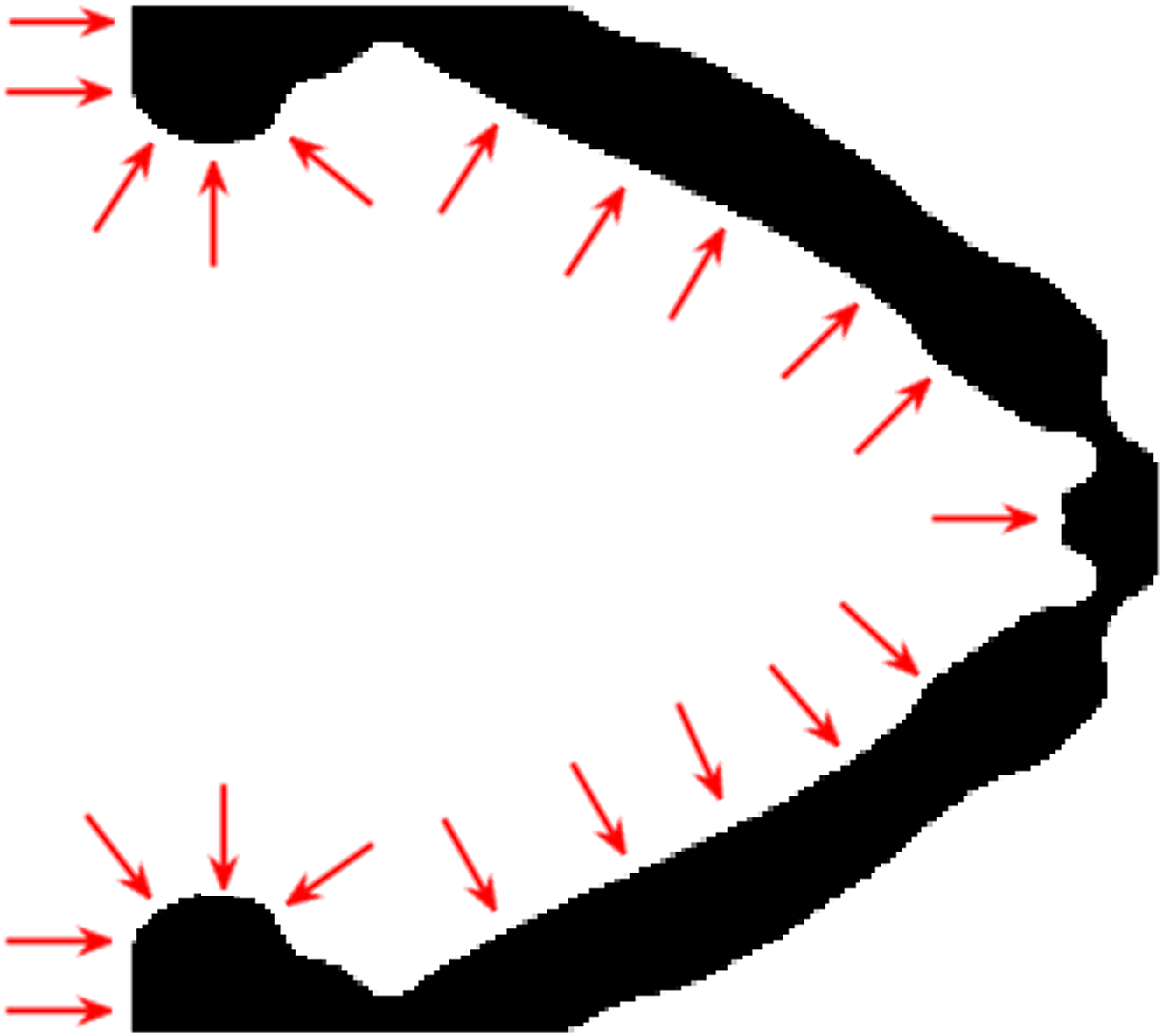} \\ \hline
\end{tabular}
\end{table}
\begin{figure}[]
  \centering  
  \begin{subfigure}[b]{0.45\textwidth} 
    \centering
    \begin{tikzpicture}[scale=0.7]
      \begin{axis}[
      xlabel={\large Iteration},
      ylabel={\large Objective function $(f_0)$},
      xmin=0,xmax=400,
      grid=major,
  ]
    \addplot[smooth,{cyan}, line width=1pt, mark = none] table[x index=0, y index=1] {Q4_Robust_Inv_objb_obje_SE95.txt}; \addlegendentry{\large Q4}
    \addplot[smooth,{red}, line width=1pt, mark = none] table[x index=0, y index=1] {Q8_Robust_Inv_objb_obje_SE95.txt}; \addlegendentry{\large Q8}
    \addplot[smooth,{green}, line width=1pt, mark = none] table[x index=0, y index=1] {Q9_Robust_Inv_objb_obje_SE95.txt}; \addlegendentry{\large Q9}
      \end{axis}
    \end{tikzpicture}
    \caption{$S_e/S^*=0.95$}
    \label{Convergence_Inverter_blueprint_95}
  \end{subfigure}
  \hfill
  \begin{subfigure}[b]{0.45\textwidth} 
    \centering
    \begin{tikzpicture}[scale=0.7]
      \begin{axis}[
      xlabel={\large Iteration},
      ylabel={\large Objective function $(f_0)$},
      xmin=0,xmax=400,
      grid=major,
  ]
    \addplot[smooth,{cyan}, line width=1pt, mark = none] table[x index=0, y index=1] {Q4_Robust_Inv_objb_obje_SE80.txt}; \addlegendentry{\large Q4}
    \addplot[smooth,{red}, line width=1pt, mark = none] table[x index=0, y index=1] {Q8_Robust_Inv_objb_obje_SE80.txt}; \addlegendentry{\large Q8}
    \addplot[smooth,{green}, line width=1pt, mark = none] table[x index=0, y index=1] {Q9_Robust_Inv_objb_obje_SE80.txt}; \addlegendentry{\large Q9}
      \end{axis}
    \end{tikzpicture}
    \caption{$S_e/S^*=0.80$}
    \label{Convergence_Inverter_blueprint_80}
  \end{subfigure}
  \caption{Convergence graphs of inverter blueprint designs}
  \label{Convergence_Inverter_blueprint}
\end{figure}
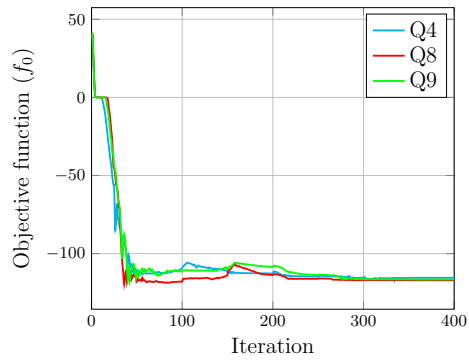
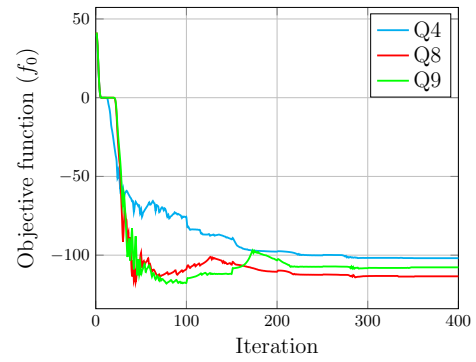

Figure \ref{Convergence_Inverter_blueprint} depicts the convergence plots for the objective function $f_0$ with respect to the iteration numbers. For strain energy fractions of 0.95 and 0.80, the graphs are shown in Figure \ref{Convergence_Inverter_blueprint_95} and Figure \ref{Convergence_Inverter_blueprint_80}, respectively. These plots demonstrate a convergent behaviour of the objective $f_0$.

\begin{table}[h!]
\centering
\caption{Optimized design for the blueprint and eroded parts (gripper)}
\label{Gripper_table}
\begin{tabular}{|C{1cm}|C{1.6cm}|C{1cm}|C{4cm}|C{4cm}|}
\hline
\textbf{QE} & $\bm{f_0}$ & $\bm{V/V^*}$ & \textbf{Blueprint} & \textbf{Eroded} \\ \hline
Q4 & \shortstack{$-160.2263$ \\ $-167.4651$} & \shortstack{$0.301$ \\ $0.276$} & \rule{0pt}{3.3cm} \includegraphics[scale=0.33]{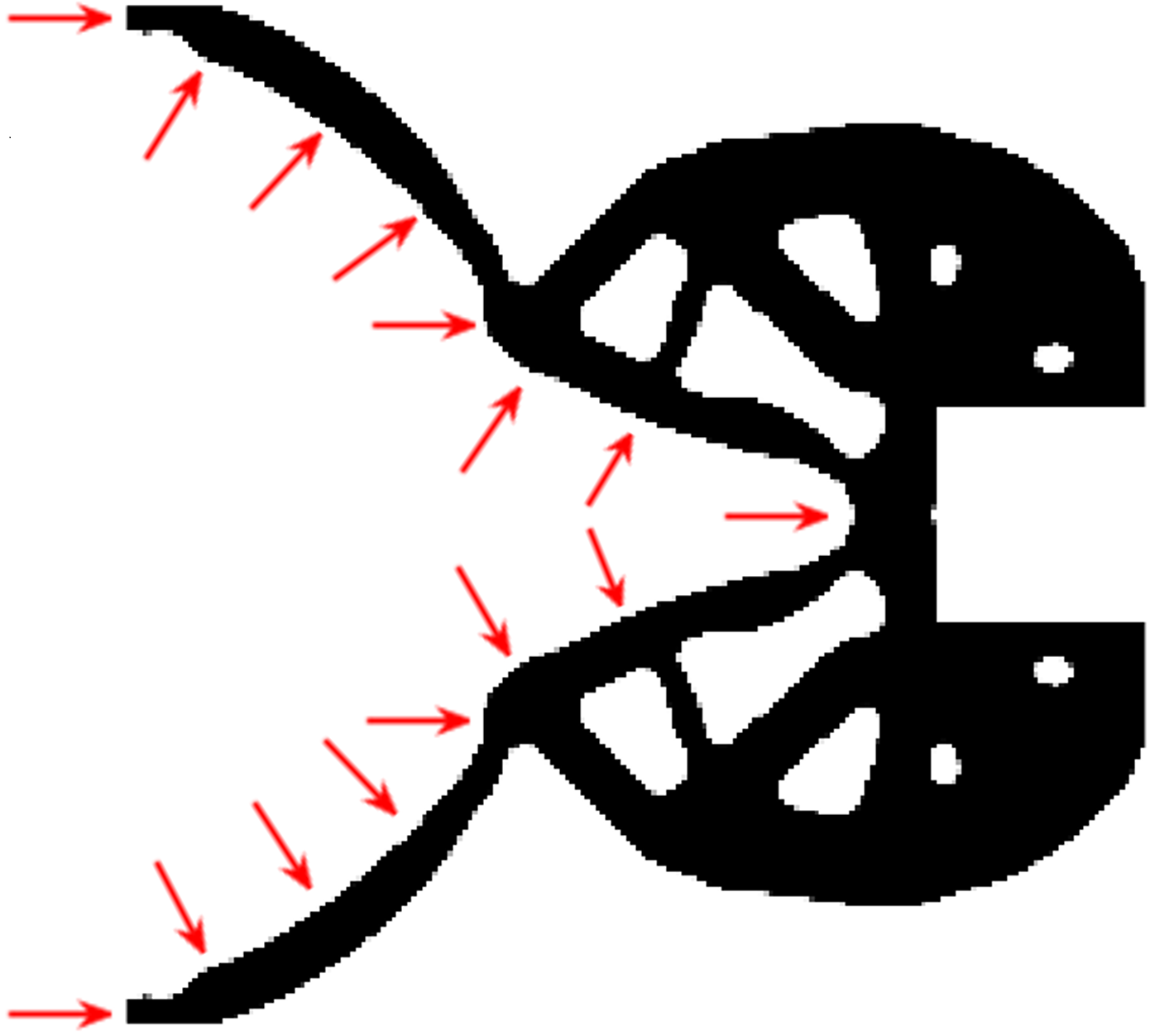} & \rule{0pt}{3.3cm} \includegraphics[scale=0.33]{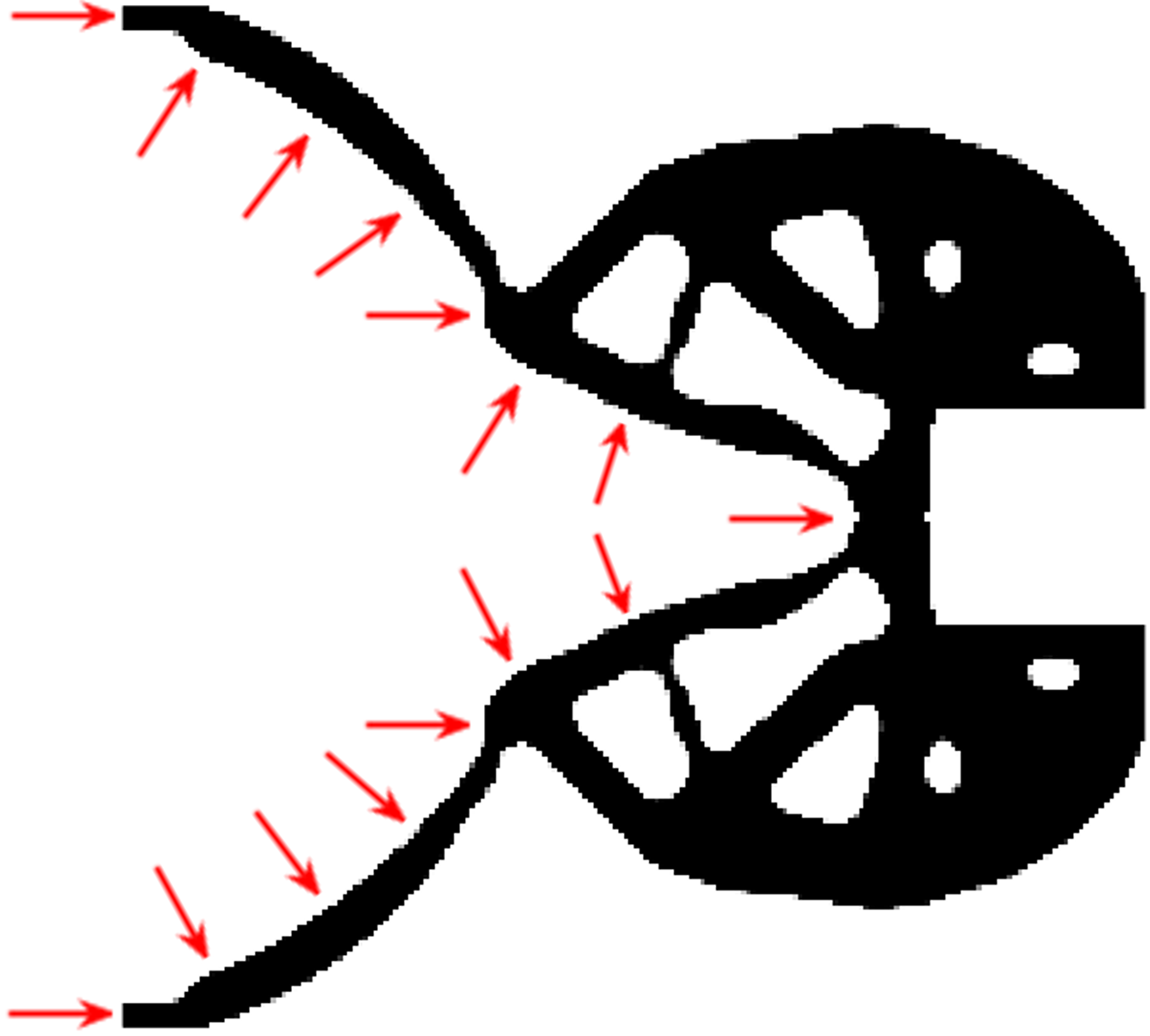} \\ \hline
Q8 & \shortstack{$-120.1879$ \\ $-132.0414$} & \shortstack{$0.299$ \\ $0.274$} & 
\rule{0pt}{3.3cm} \includegraphics[scale=0.33]{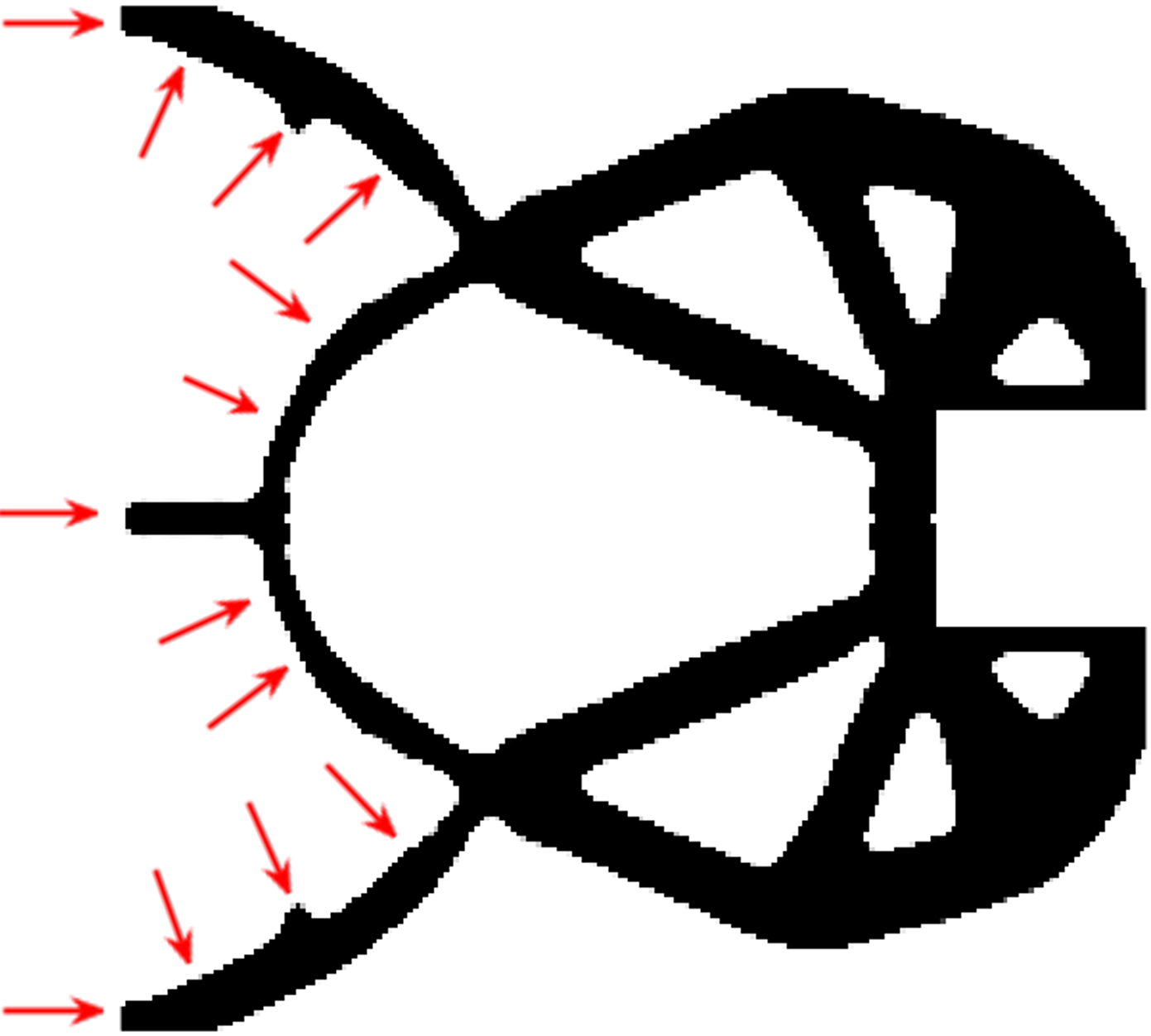} &
\rule{0pt}{3.3cm} \includegraphics[scale=0.33]{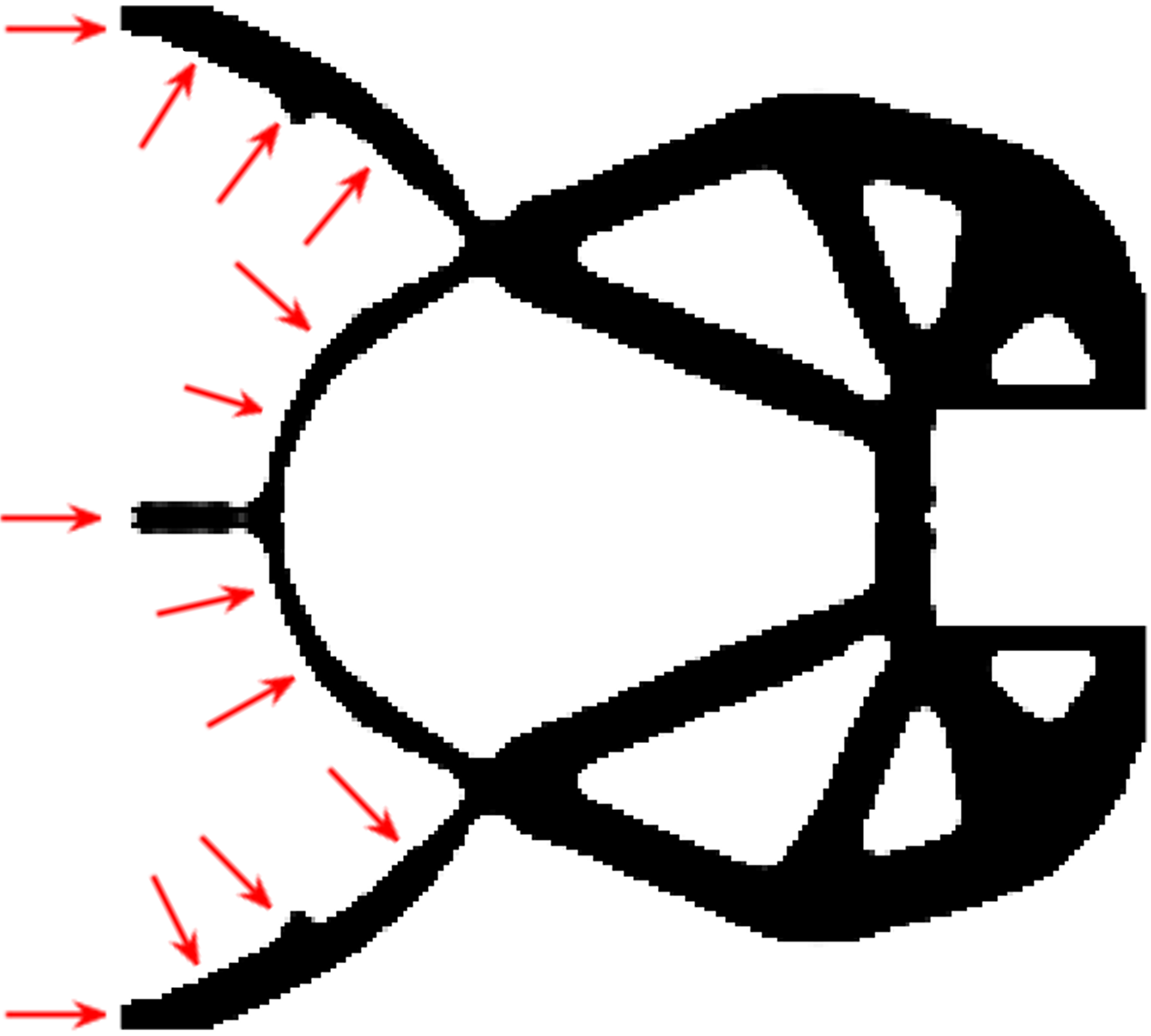} \\ \hline
Q9 & \shortstack{$-116.6134$ \\ $-127.2994$} & \shortstack{$0.299$ \\ $0.274$} & \rule{0pt}{3.3cm} \includegraphics[scale=0.33]{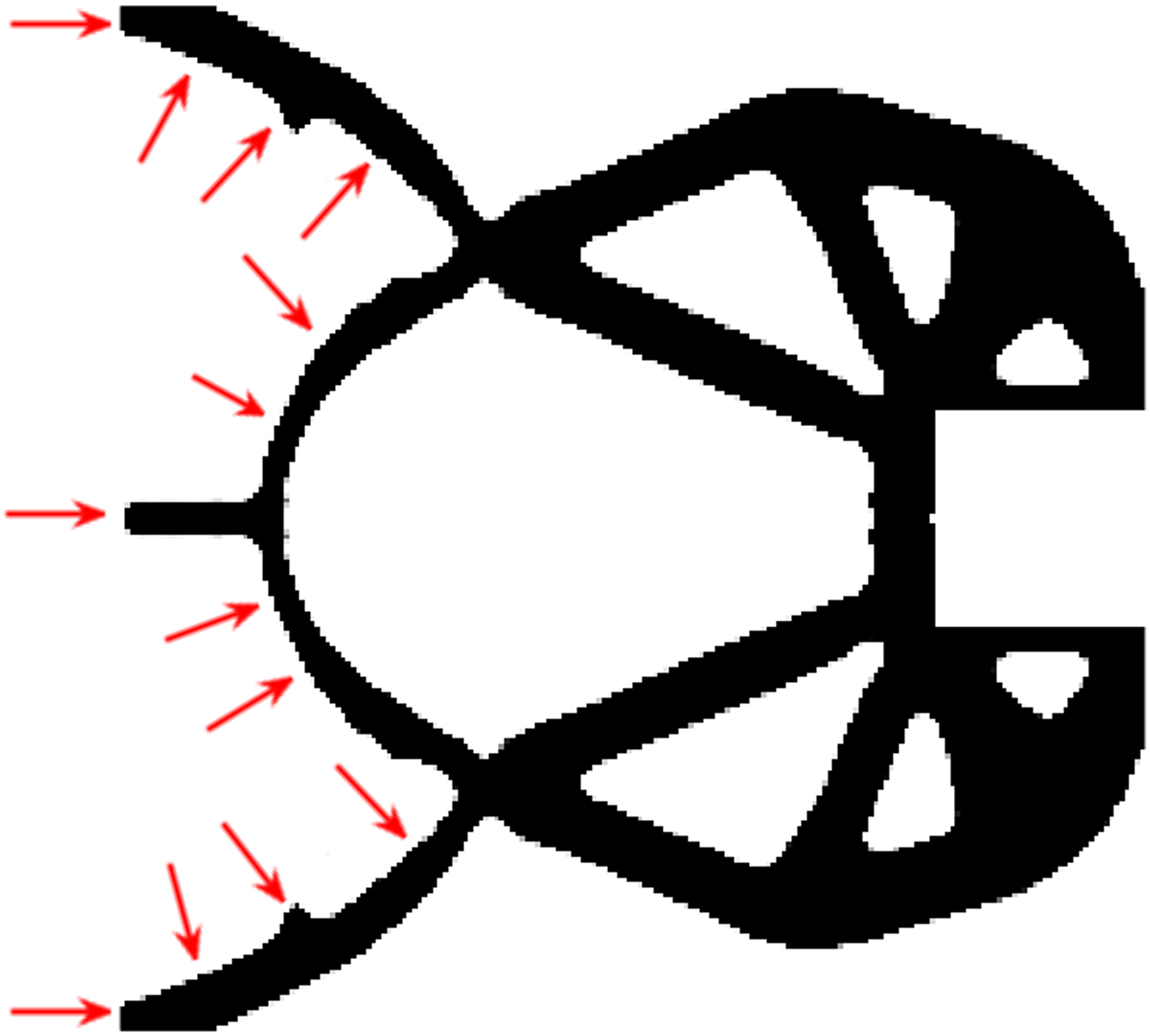} & \rule{0pt}{3.3cm} \includegraphics[scale=0.33]{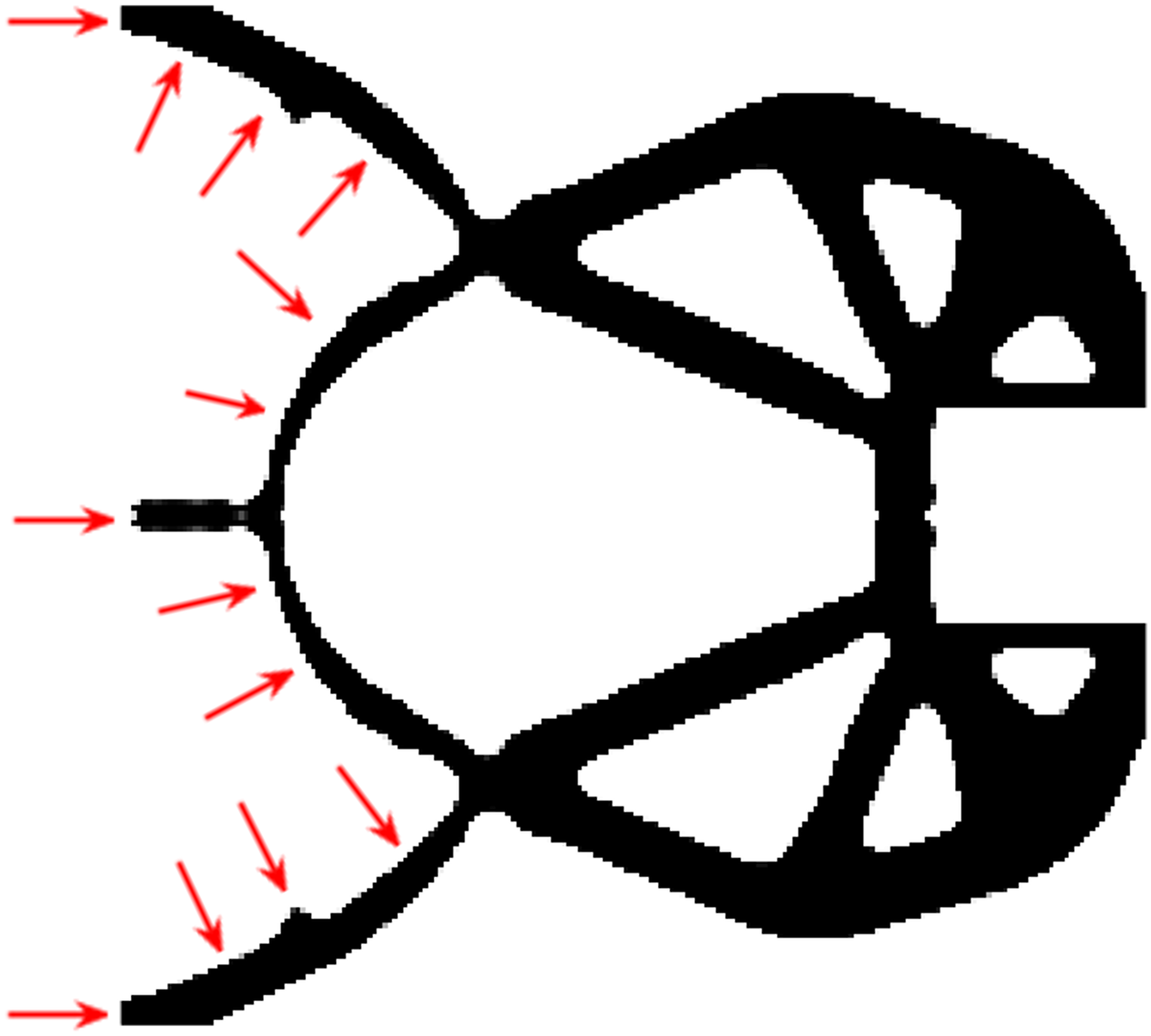} \\ \hline
\end{tabular}
\end{table}

Table \ref{Gripper_table} shows all the optimized results obtained for the gripper. Q4 blueprint and eroded results have better $f_0$ than Q8 and Q9. The optimized designs are topologically different. 
The topology of Q4 results is much different from the others. Additionally, the value of $f_0$ is less for eroded designs, i.e., eroded designs give better output displacement than blueprints. Furthermore, the volume fraction achieved in optimized results for eroded designs is less than that of blueprint designs.

In all of the above cases, the magnitude of the objective function is higher for the eroded than the blueprint mechanisms (i.e., more flexible). As the eroded design has less volume and less strain energy compared to the blueprint, it tends to be more compliant. However, for all purposes, blueprint mechanisms should be used as they are relatively more practical to realize. 

\section{Conclusions}\label{sec6}
This work compares the performance of optimized design-dependent pressure-actuated compliant mechanisms obtained using standard and higher-order quadrilateral meshing elements (Q4, Q8, and Q9). An inverter and a gripper mechanism are taken for the study. The robust formulation is used, and the blueprint and eroded designs are considered. The min-max optimization formulation using the output deformation of these designs is formulated. A volume constraint and a strain energy constraint are applied to the blueprint and eroded designs, respectively. Q8 suggests more output displacement for inverters, whereas Q4 yields a better value of the objective function for the gripper. A greater magnitude of the objective function indicates higher flexibility of the design. Using standard and higher-order quadrilateral meshing elements reveals a slight variation in optimized results, indicating that the results are element-dependent. The eroded design of the gripper mechanism provides more flexibility at the output than its blueprint design. The study demonstrates that using Q4, Q8 and Q9 quadrilateral finite elements impacts both the resulting topologies and performance outcomes.


\end{document}